\documentclass[twocolumn,aps,showpacs]{revtex4}

\usepackage{graphicx}% Include figure files
\usepackage{dcolumn}% Align table columns on decimal point
\usepackage{bm}% bold math

\begin{document}

\title{Local entanglement of multidimensional continuous-variable systems}

\author{H.-C. Lin}

\email{ho.lin@ucl.ac.uk}

\author{A. J. Fisher}

\email{andrew.fisher@ucl.ac.uk}

\affiliation{UCL Department of Physics and Astronomy and London Centre for Nanotechnology,\\
 University College London, Gower Street, London WC1E 6BT, U.K.}

\begin{abstract}
We study the `local entanglement' remaining after filtering operations corresponding to imperfect measurements performed by one or both parties, such that the parties can only determine whether or not the system is located in some region of space. The local entanglement in pure states of general bipartite multidimensional continuous-variable systems can be completely determined  through simple expressions. We apply our approach to semiclassical WKB systems, multi-dimensional harmonic oscillators, and a hydrogen atom as three examples.
\end{abstract}


\maketitle

\section{Introduction}\label{sec:intro}
It has been recognized quite recently that quantum entanglement is not just a profound feature of quantum mechanics but it is also a valuable physical resource, like energy, with massive potential for technological applications, such as quantum computation \cite{Nielsenbook}, quantum cryptography \cite{Ekert1991} and quantum teleportation \cite{Bennett1993}, etc. However, our understanding of entanglement is still far from complete despite current intense research activities.

There are many reasons to focus on the entanglement of continuous-variable states \cite{ Braunstein.RevModPhys.77.531,Braunsteinbook,Eisert2003}, since the underlying degrees of freedom of physical systems carrying quantum information are frequently continuous, rather than discrete. Much of the effort has been concentrated on Gaussian states (i.e., states whose Wigner function is a Gaussian), since these are common (especially in quantum optics) as the ground or thermal states of optical modes. Within this framework, many interesting topics have been studied; for example, entanglement distillation for Gaussian states \cite{Duan2000,Eisert2002,Fiurasek2002,Giedke2002}, multipartite entangled Gaussian states \cite{Loock2000,Giedke2001,Adesso2004} and entanglement measures, such as entanglement of formation \cite{giedke:107901,Wolf2004}  and logarithmic negativity \cite{PhysRevA.65.032314,PhysRevA.66.042327}. However one should remember that non-Gaussian states are also extremely important; this is especially so in condensed-phase systems, where harmonic behavior in any degree of freedom is likely to be only an approximation.  Much less is known about the entanglement of these non-Gaussian states: while there is some progress in finding criteria for entanglement \cite{shchukin05},  there is little knowledge about how to quantify it.

In two preceding papers \cite{lin1,lin2}, we demonstrated how
to use a specific type of projective filtering to characterize the
distribution (particularly in configuration space) of entanglement
in any smooth two-mode bipartite continuous-variable state.  The approach is based on making an imperfect measurement of the `position' of the system in configuration space, and then studying the entanglement remaining after the measurement.  We showed how this approach could be used to map entanglement in different situations \cite{lin1}, and that simple formulae exist for the entanglement in the limit where the region to which the system is confined after the measurement becomes small (i.e., where the measurement becomes more and more accurate) \cite{lin2}.

In this paper we generalize these results to general (including multimode) smooth bipartite pure states.  We first review the important results for two-mode states in \S\ref{sec:two-mode}, then generalize to multi-mode states in \S\ref{sec:multi-mode}.  Finally in \S\ref{sec:examples} we show examples of our approach applied to some systems in which analytical expressions for the energy eigenfunctions are easily obtained, before giving our conclusions in \S\ref{sec:conclusions}.

\section{Two-mode states}\label{sec:two-mode}
We briefly recapitulate the definitions
of essential terms and the known results for any smooth bipartite
two-mode continuous-variable state. Let Alice and Bob share a state
of two distinguishable one-dimensional particles. Alice can measure
only the position of her particle (coordinate $q_{A}$), Bob the position
of his (coordinate $q_{B}$). They filter their state by determining
whether or not the particles are found in particular regions of configuration
space, and discard instances in which they are not. We refer to the
resulting subensemble as the ``discarding ensemble''. On the
other hand if they choose not to discard the system when the particles
are not in the desired regions, the resulting subensemble is called
the ``nondiscarding ensemble''. The entanglement $\mathcal{E}_{D}$
in the discarding ensemble is related to the entanglement $\mathcal{E}_{ND}$
in the nondiscarding ensemble by 
\begin{equation}\label{eq:nondiscdef}
\mathcal{E}_{ND}=p_{ab}\mathcal{E}_{D},
\end{equation}
where $p_{ab}$ is the probability of finding Alice's particle within
the region $\{ q_{A}:\bar{q}_{A}-a\le q_{A}\le\bar{q}_{A}+a\}$ and
Bob's particle in the region $\{ q_{B}:\bar{q}_{B}-b\le q_{B}\le\bar{q}_{B}+b\}$.
We shall therefore focus on calculating $\mathcal{E}_{D}$, noting
that $\mathcal{E}_{ND}$ can be simply obtained from it; we show plots for both quantities for some of the systems discussed in \S\ref{sec:examples}.

\subsection{Preliminary measurements on Alice's particle only}\label{sec:two-mode-Alice-only}
If the initial state  $\hat{\rho}$ is pure, so is $\hat{\rho}_{D}$ in the discarding ensemble. 
Suppose the initial filtering is performed only by Alice, by determining whether
$q_{A}$ lies in the region $\bar{q}_{A}-a\le q_{A}\le\bar{q}_{A}+a$,
and all instances in which this is not the case are discarded. Now,
since $a$ is to be very small, Alice's original (before the measurement) reduced density matrix
$\rho^{(A)}$($=\mathrm{Tr}_B[\hat{\rho}]$) in the neighborhood of
$\bar{q}_{A}$ can be expanded (provided it is smooth in configuration
space) as 
\begin{eqnarray}\label{eq:rhoaexpand}
 &  & \rho^{(A)}(q_{A};q'_{A})=\rho^{(A)}(\bar{q}_{A}+x;\bar{q}_{A}+x')\\
 &  & \quad=\rho_{00}^{(A)}+\rho_{10}^{(A)}x+\rho_{01}^{(A)}x'+\rho_{11}^{(A)}xx'+\mathrm{O}(x^{2},x'^{2}),\nonumber \end{eqnarray}
 where \begin{equation}
\rho_{n_{1}n_{2}}^{(A)}=\frac{\partial^{n_{1}}}{\partial{q_{A}}^{n_{1}}}\frac{\partial^{n_{2}}}{\partial{q'_{A}}^{n_{2}}}\rho^{(A)}(q_{A},q'_{A})\Big\vert_{q_{A}=q'_{A}=\bar{q}_{A}}.\end{equation}
Within region $-a\le x\le a$, $\rho_{D}^{(A)}$ is obtained by dividing equation~(\ref{eq:rhoaexpand}) by the normalizing factor  $2a\rho_{00}^{(A)}+\mathrm{O}(a^3)$.

Now seek right eigenfunctions $\phi_n$ of $\rho_{D}^{(A)}$ within the allowed region:
\begin{equation}\label{eq:eigcondition}
\int_{-a}^a\mathrm{d} x'\,\rho^{(A)}(x;x')\phi_n(x')=\lambda_n\phi_n(x).
\end{equation}
Expanding $\phi_n$ as a power series
\begin{equation}
\phi_n(x)=a_n+b_n(x)+{c_n\over 2}x^2+\mathrm{O}(x^3),\label{eq:phiexpand}
\end{equation}
the eigenfunction condition becomes (to order $a^3$)
\begin{widetext}\begin{equation}
{1\over 2a[\rho^{(A)}_{00}+\mathrm{O}(a^2)]}\left[
a\left(\begin{array}{ccc}
2\rho^{(A)}_{00}&0&\ldots\\
2\rho^{(A)}_{10}&0&\ldots\\
\vdots&\vdots&\ddots
\end{array}\right)
+a^3\left(\begin{array}{ccc}
\rho^{(A)}_{20}/3&2\rho^{(A)}_{01}/3&\ldots\\
0&2\rho^{(A)}_{11}/3&\ldots\\
\vdots&\vdots&\ddots
\end{array}\right)
\right]
\left(\begin{array}{c}a_n\\b_n\\\vdots\end{array}\right)
\equiv{\bf M}\left(\begin{array}{c}a_n\\b_n\\\vdots\end{array}\right)
=\lambda_n\left(\begin{array}{c}a_n\\b_n\\\vdots\end{array}\right).
\end{equation}\end{widetext}
Expanding $\det(M-\lambda I)$ to order $a^4$ and equating to zero, 
we find two non-zero eigenvalues: \begin{eqnarray}
\lambda_{1} & = & \frac{a^{2}}{3{\rho^{(A)}_{00}}^{2}}(\rho_{11}^{(A)}\rho_{00}^{(A)}-\rho_{01}^{(A)}\rho_{10}^{(A)})\nonumber \\
\lambda_{2} & = & 1-\lambda_{1}.\label{eq:2Dlambdas}
\end{eqnarray}
So to the lowest non-trivial order ($a^2$), the eigenvalues, and hence the von
Neumann entropy, of $\rho_{D}^{(A)}$ are entirely determined by the
quantity $\epsilon\equiv\lambda_{1}$. Specifically, the von Neumann
entropy is 
\begin{equation}\label{eq:vonNeuman}
S_v=h(\epsilon)\equiv-[\epsilon\log_{2}(\epsilon)+(1-\epsilon)\log_{2}(1-\epsilon)].\end{equation}

To find the leading corrections to this result, we include all terms proportional to $x^2$ or $x'^2$ in the expansion (\ref{eq:rhoaexpand}) for $\rho^{(A)}$: 
\begin{eqnarray}
\rho^{(A)}(x;x')&=&\rho^{(A)}_{00}+\rho^{(A)}_{10}x+\rho^{(A)}_{01}x'\\
&  &\quad+\frac{1}{2}(\rho^{(A)}_{20}x^2+\rho^{(A)}_{02}x'^2+2\rho^{(A)}_{11}xx')\nonumber\\
&  &\quad+\frac{1}{2}(\rho^{(A)}_{21}x^2 x'+\rho^{(A)}_{12}x x'^2)\nonumber \\
&  &\quad+\frac{1}{4} \rho^{(A)}_{22} x^2 x'^2+\mathrm{O}(x^3,x'^3).\nonumber
\end{eqnarray}
and then carry equation~(\ref{eq:phiexpand}) to third order:
\begin{equation}
\phi_n(x)=a_n+b_n x+\frac{1}{2}c_n x^2+\frac{1}{6}d_n x^3+\mathrm{O}(x^4),
\end{equation}
From the eigenfunction condition (\ref{eq:eigcondition}), we find the third non-zero eigenvalue to be
\begin{eqnarray}\label{eq:AliceOnlyThirdEigval}
\lambda_3&=&\frac{a^4}{ 90{\rho^{(A)}_{00}}^{2} (\rho^{(A)}_{01}\rho^{(A)}_{10}-\rho^{(A)}_{11}\rho^{(A)}_{00})}\left( \rho^{(A)}_{02}\rho^{(A)}_{11}\rho^{(A)}_{20}\right. \nonumber\\
 &  &+\rho^{(A)}_{01}\rho^{(A)}_{22}\rho^{(A)}_{10}+\rho^{(A)}_{12}\rho^{(A)}_{00}\rho^{(A)}_{21}-\rho^{(A)}_{01}\rho^{(A)}_{12}\rho^{(A)}_{20}\nonumber\\
 &  &\left. -\rho^{(A)}_{10}\rho^{(A)}_{02}\rho^{(A)}_{21}-\rho^{(A)}_{00}\rho^{(A)}_{11}\rho^{(A)}_{22}\right) +\mathrm{O}(a^6).\end{eqnarray}
Therefore, the corrections due to higher eigenvalues, arising from the higher-order terms in equation~(\ref{eq:rhoaexpand}),
affect $\epsilon$ (and hence the entanglement) only to order $a^{4}$.

\subsection{Preliminary measurements on both particles}
Now suppose both parties restrict their measurements: Alice's particle
must lie in $\{ q_{A}:\bar{q}_{A}-a\le q_{A}\le\bar{q}_{A}+a\}$,
and Bob's in $\{ q_{B}:\bar{q}_{B}-b\le q_{B}\le\bar{q}_{B}+b\}$.  In \cite{lin2} we attacked this problem by reducing it to an effective two-qubit one, for which exact results are available.  However this approach does not generalize so naturally to the multi-mode case, so we give here an alternative approach. From the argument above we know we can compute the entanglement from
Alice's reduced density matrix $\rho^{(A)}$ in the coordinate representation.
Our first task, therefore, is to evaluate this quantity once Bob has
made the measurement of his particle.

We do this by making a further Taylor expansion involving Bob's variables.  We define 
\begin{eqnarray}
\rho_{n_{1}n_{2}n_{3}n_{4}} & = & \frac{\partial^{n_{1}}}{\partial{q_{A}}^{n_{1}}}\frac{\partial^{n_{2}}}{\partial{q'_{A}}^{n_{2}}}\frac{\partial^{n_{3}}}{\partial{q_{B}}^{n_{3}}}\frac{\partial^{n_{4}}}{\partial{q'_{B}}^{n_{4}}}\nonumber \\
&   &\quad\rho(q_{A},q_{B};q'_{A},q'_{B})\Big\vert_{\bar{q}_{A},\bar{q}_{B}}.\end{eqnarray}
As we will see, to obtain the first nontrivial term in the solution
we need all terms to first order in Alice's coordinates and to second
order in Bob's: \begin{eqnarray}
 &  &\rho(q_{A},q_{B};q'_{A},q'_{B})\nonumber \\
 & = & \rho(\bar{q}_{A}+x_{A},\bar{q}_{B}+x_{B};\bar{q}_{A}+x'_{A},\bar{q}_{B}+x'_{B})\nonumber \\
 & = & \rho_{0000}+\rho_{1000}x_{A}+\rho_{0100}x'_{A}+\rho_{0010}x_{B}+\rho_{0001}x'_{B}\nonumber \\
 &  & \quad+\frac{1}{2}(\rho_{0020}{x_{B}}^{2}+\rho_{0002}{x'_{B}}^{2})\nonumber \\
 &  & \quad+\rho_{1100}x_{A}x'_{A}+\rho_{1010}x_{A}x_{B}+\rho_{1001}x_{A}x'_{B}\nonumber \\
 &  & \quad+\rho_{0110}x'_{A}x_{B}+\rho_{0101}x'_{A}x'_{B}+\rho_{0011}x_{B}x'_{B}\nonumber \\
 &  & \quad+\frac{1}{2}(\rho_{1020}x_{A}{x_{B}}^{2}+2\rho_{1011}x_{A}x_{B}x'_{B}+\rho_{1002}x_{A}{x'_{B}}^{2}\nonumber \\
 &  & \qquad+\rho_{0120}x'_{A}{x_{B}}^{2}+2\rho_{0111}x_{A'}x_{B}x'_{B}+\rho_{0102}x'_{A}{x'_{B}}^{2})\nonumber \\
 &  & \quad+\mathrm{O}({x_{A}}^{2},{x'_{A}}^{2},{x_{B}}^{3},{x'_{B}}^{3}).\end{eqnarray}

Alice's reduced density matrix is then found by writing 
\begin{eqnarray}
\rho^{(A)}(x_{A};x'_{A}) & = & \frac{1}{p}\int_{-b}^{b}\mathrm{d} x_{B}\,\rho(x_{A},x_{B};x'_{A},x_{B})\nonumber \\
 & = & \frac{2b}{p}[\rho_{0000}+x_{A}\rho_{1000}+x'_{A}\rho_{0100}]\nonumber \\
 &  & \quad+\frac{b^{3}}{3}[\rho_{0020}+2\rho_{0011}+\rho_{0002}\nonumber \\
 &  & \quad+(\rho_{1020}+2\rho_{1011}+\rho_{1002})x_{A}\nonumber \\
 &  & \quad+(\rho_{0120}+2\rho_{0111}+\rho_{0102})x'_{A}]\nonumber \\
 &  & \quad+\mathrm{O}(b^{5},{x_{A}}^{2},{x'_{A}}^{2}).
 \end{eqnarray}
where $p$ is a normalization constant. By comparison with equation (\ref{eq:rhoaexpand}) and equating powers
of $x_{1}$ and $y_{1}$ we can immediately identify the terms which
appear in the expression for $\epsilon$,
and therefore determine the entanglement:
\begin{eqnarray}
\rho^{(A)}_{00} & = & \frac{2b}{p}[\rho_{0000}+\frac{b^{2}}{6}(\rho_{0020}+2\rho_{0011}+\rho_{0002})]+\mathrm{O}(b^{5});\nonumber \\
\rho^{(A)}_{10} & = & \frac{2b}{p}[\rho_{1000}+\frac{b^{2}}{6}(\rho_{1020}+2\rho_{1011}+\rho_{1002})]+\mathrm{O}(b^{5});\nonumber \\
\rho^{(A)}_{01} & = &\frac{2b}{p}[\rho_{0100}+\frac{b^{2}}{6}(\rho_{0120}+2\rho_{0111}+\rho_{0102})]+\mathrm{O}(b^{5});\nonumber \\
\rho^{(A)}_{11} & = & \frac{2b}{p}[\rho_{1100}+\frac{b^{2}}{6}(\rho_{1120}+2\rho_{1111}+\rho_{1102})]+\mathrm{O}(b^{5}).\nonumber \\
&  & 
\label{eq:termsforentanglement}
\end{eqnarray}

The leading (order $b^{2}$) terms in the numerator of the expression
for $\epsilon$ cancel---this
is the reason why we need the density matrix to quadratic order in
Bob's coordinates. The cancellation occurs because Alice and Bob (by
hypothesis) share a pure state, and so 
\begin{eqnarray}\label{eq:rhopsiderivs}
&&\rho(q_A,q_B;q'_A,q'_B)=\psi(q_A,q_B)\psi^*(q'_A,q'_B)\nonumber\\
&&\quad\Rightarrow\rho_{n_{1}n_{2}n_{3}n_{4}}={\partial^{n_{1}}\over\partial {q_A}^{n_{1}}}{\partial^{n_{3}}\over\partial {q_B}^{n_{3}}}\psi(q_A,q_B)\Big\vert_{\bar{q}_A,\bar{q}_B}\nonumber\\
&&\qquad\times
{\partial^{n_{2}}\over\partial {q'_A}^{n_{2}}}{\partial^{n_{4}}\over\partial {q'_B}^{n_{4}}}\psi^*(q'_A,q'_B)\Big\vert_{\bar{q}_A,\bar{q}_B}.
\end{eqnarray}
We can thus re-arrange the indices in a product of two $\rho_{n_{1}n_{2}n_{3}n_{4}}$
terms as \begin{equation}
\rho_{abcd}\rho_{efgh}=\rho_{ebgd}\rho_{afch},\label{eq:rearrange}\end{equation}
 so in particular \begin{equation}
\rho_{1100}\rho_{0000}=\rho_{0100}\rho_{1000}.\end{equation}

Hence the leading term in the numerator of $\epsilon$
is of order $b^{4}$, and the overall expression becomes \begin{eqnarray}
\epsilon & = & \frac{a^{2}b^{2}}{18\rho_{0000}^{2}}[\rho_{1100}(\rho_{0020}+2\rho_{0011}+\rho_{0002})\nonumber \\
 &  & \quad+\rho_{0000}(\rho_{1120}+2\rho_{1111}+\rho_{1102})\nonumber \\
 &  & \quad-\rho_{1000}(\rho_{0120}+2\rho_{0111}+\rho_{0102})\nonumber \\
 &  & \quad-\rho_{0100}(\rho_{1020}+2\rho_{1011}+\rho_{1002})].\end{eqnarray}
 Using equation (\ref{eq:rearrange}) we can simplify this to obtain 
\begin{eqnarray}
\frac{\epsilon}{a^{2}b^{2}} & = & \frac{1}{9\rho_{0000}^{2}}[\rho_{1100}\rho_{0011}+\rho_{0000}\rho_{1111}\nonumber \\
 &  & \quad-\rho_{1000}\rho_{0111}-\rho_{0100}\rho_{1011}]\label{eq:firstformresult}\\
 & = & \frac{1}{18\rho_{0000}^{2}}[2\rho_{1100}\rho_{0011}+2\rho_{0000}\rho_{1111}-\rho_{1000}\rho_{0111}\nonumber \\
 &  & \quad-\rho_{0100}\rho_{1011}-\rho_{0010}\rho_{1101}-\rho_{0001}\rho_{1110}].\label{eq:secondformresult}\end{eqnarray}
The first form (\ref{eq:firstformresult}) is slightly more compact,
while the second form (\ref{eq:secondformresult}) makes it clear
that the coordinates of Alice's and Bob's subsystems are treated equivalently,
as required.  The von Neumann entropy, and hence the entanglement (since this is
still a pure state), is then $S_v=h(\epsilon)$ as before.

We know, from the arguments leading to equation~(\ref{eq:AliceOnlyThirdEigval}), that the leading correction to this result is $\mathrm{O}(a^4)$, and we should expect from the symmetry between Alice's and Bob's systems that it is also $\mathrm{O}(b^4)$.  We have explicitly computed the correction and this is indeed the case: the result is given in Appendix~\ref{app:a4b4correction}. The third eigenvalue $\lambda_3$ measures the extent of the breakdown of our approach.  We note that it is of order $a^{4}b^{4}$, and therefore does not affect the expression of $\epsilon$, which is of order $a^{2}b^{2}$.

\section{Multi-dimensional systems}\label{sec:multi-mode}

\subsection{General approach}
Consider first the case in which only Alice makes preliminary measurements.  If Alice's system is two-dimensional and she localizes the particle so $-a_i\le x_i\le +a_i, i\in \{1,2\}$, one can find the eigenvalues of $\rho^{(A)}$ by a straightforward generalization of the methods in \S\ref{sec:two-mode-Alice-only}.  Once again we find that there are only two non-zero eigenvalues to order $a_i^2$:
\begin{eqnarray}\label{eq:twodim}
\lambda_{1} & = & \sum_{i}^{2}\frac{a_{i}^{2}}{3(\bar{\rho}^{(A)})^{2}}\left(\bar{\rho}^{(A)}\frac{\partial^{2}\rho^{(A)}}{\partial q_{A,i}\partial q'_{A,i}}-\frac{\partial\rho^{(A)}}{\partial q_{A,i}}\frac{\partial\rho^{(A)}}{\partial q'_{A,i}}\right)\nonumber \\
&  & \quad+\mathrm{H.T.}\nonumber \\
\lambda_{2} & = & 1-\lambda_{1}.
\end{eqnarray}
where $i$ goes over the two spatial dimensions of Alice's subsystems, H.T. stands for higher-order terms and $\bar{\rho}^{(A)}=\rho^{(A)}(\mathbf{\bar{q}}_{A};\mathbf{\bar{q}}_{A})$.

We now argue that this property holds irrespective of the dimensionality of Alice's system, as follows.  The entanglement must be invariant under exchange of the axis labels, and under all transformations of the form $a_i\rightarrow -a_i$.  The only possibilities consistent with these requirements are
\begin{equation}\label{eq:firstform}
\lambda_1=1-\sum_i t_i a_i^2; \quad \lambda_2=\sum_i t_i a_i^2; \quad \lambda_3,\lambda_4\ldots=0,
\end{equation}
or
\begin{equation}\label{eq:secondform}
\lambda_1=1-\sum_i t_i a_i^2; \quad \lambda_2=t_1 a_1^2; \quad \lambda_3=t_2 a_2^2,\ldots,
\end{equation}
where the $t_i$ are arbitrary constants.
Furthermore the eigenvalues must reduce to the known forms for one- and two-dimensional systems if all other $a_i$ are set to zero.  If we keep $a_1$ and $a_2$ non-zero, sending all others to zero, only the first form (\ref{eq:firstform}) is consistent with equation (\ref{eq:twodim}).  Therefore, the form of the non-zero eigenvalues must be
\begin{eqnarray}
\lambda_{1} & = & \sum_{i}\frac{a_{i}^{2}}{3(\bar{\rho}^{(A)})^{2}}\left(\bar{\rho}^{(A)}\frac{\partial^{2}\rho^{(A)}}{\partial q_{A,i}\partial q'_{A,i}}-\frac{\partial\rho^{(A)}}{\partial q_{A,i}}\frac{\partial\rho^{(A)}}{\partial q'_{A,i}}\right)\nonumber \\
&   & \quad+\mathrm{H.T.}\nonumber \\
\lambda_{2} & = & 1-\lambda_{1}\label{eq:multiDlambdas}\\
\lambda_{3} & = & 0+\mathrm{H.T.}\nonumber \end{eqnarray}
where $i$ now goes over all the dimensions of Alice's subsystems.

Define
\begin{eqnarray}
\rho_{(i,j;n_{1}n_{2}n_{3}n_{4})} & = & {\frac{\partial^{n_{1}}}{\partial{q_{A,i}}^{n_{1}}}}{\frac{\partial^{n_{2}}}{\partial{q'_{A,i}}^{n_{2}}}}{\frac{\partial^{n_{3}}}{\partial{q_{B,j}}^{n_{3}}}}{\frac{\partial^{n_{4}}}{\partial{q'_{B,j}}^{n_{4}}}}\nonumber \\
& & \quad\rho(\mathbf{q}_{A},\mathbf{q}'_{A},\mathbf{q}_{B},\mathbf{q}'_{B})\big\vert_{\bar{q}_{A,i},\bar{q}_{B,j}}.\end{eqnarray}
where $i$ ($j$) represents one of available dimensions of Alice's
(Bob's) subsystem. If the state $\rho(\mathbf{q}_{A},\mathbf{q}'_{A},\mathbf{q}_{B},\mathbf{q}'_{B})$
is pure, we have the following relation
\begin{equation}
\rho_{(i,j;n_{1}n_{2}n_{3}n_{4})}\rho_{(i,j;n_{5}n_{6}n_{7}n_{8})}=\rho_{(i,j;n_{5}n_{2}n_{7}n_{4})}\rho_{(i,j;n_{1}n_{6}n_{3}n_{8})}.\label{eq:pureRhoRelation}\end{equation}

From the previous analysis that led to equation~(\ref{eq:termsforentanglement}) for a pure two-mode state, we know we can extend
equation  (\ref{eq:multiDlambdas}) to a pure  multi-dimensional
bipartite state $\rho(\mathbf{q}_{A},\mathbf{q}'_{A},\mathbf{q}_{B},\mathbf{q}'_{B})$
for the case where both parties make preliminary measurements on their
particles by making the following substitutions:
\begin{widetext}\begin{eqnarray}
\rho^{(A)}_{00} & = & \sum_{j}\left(\frac{\prod_{j'}2b_{j'}}{p}\right)\left[ \rho_{(ij;0000)}+\frac{b_{j}^{2}}{6}(\rho_{(ij;0020)}+2\rho_{(ij;0011)}+\rho_{(ij;0002)})\right] +\mathrm{H.T.};\nonumber \\
\rho^{(A)}_{10} & = & \sum_{j}\left(\frac{\prod_{j'}2b_{j'}}{p}\right)\left[ \rho_{(ij;1000)}+\frac{b_{j}^{2}}{6}(\rho_{(ij;1020)}+2\rho_{(ij;1011)}+\rho_{(ij;1002)})\right] +\mathrm{H.T.};\nonumber \\
\rho^{(A)}_{01} & = & \sum_{j}\left(\frac{\prod_{j'}2b_{j'}}{p}\right)\left[ \rho_{(ij;0100)}+\frac{b_{j}^{2}}{6}(\rho_{(ij;0120)}+2\rho_{(ij;0111)}+\rho_{(ij;0102)})\right] +\mathrm{H.T.};\nonumber \\
\rho^{(A)}_{11} & = & \sum_{j}\left(\frac{\prod_{j'}2b_{j'}}{p}\right)\left[ \rho_{(ij;1100)}+\frac{b_{j}^{2}}{6}(\rho_{(ij;1120)}+2\rho_{(ij;1111)}+\rho_{(ij;1102)})\right] +\mathrm{H.T.}
\end{eqnarray}\end{widetext}
where $j$ and $j'$ go over all the dimensions of Bob's
subsystem and $p$ is an appropriate normalization constant.

Therefore, to
the lowest order in $a$ and $b$, $\lambda_{1}$ in equation (\ref{eq:multiDlambdas})  becomes 
\begin{eqnarray}
\lambda_{1} & = & \sum_{i,j}\frac{a_{i}^{2}b_{j}^{2}}{18{\rho_{(i,j;0000)}}^{2}}\big\lbrace \rho_{(i,j;1100)}[\rho_{(i,j;0020)}\\
 &  & \quad+2\rho_{(i,j;0011)}+\rho_{(i,j;0002)}]\nonumber \\
 &  & \quad+\rho_{(i,j;0000)}[\rho_{(i,j;1120)}+2\rho_{(i,j;1111)}+\rho_{(i,j;1102)}]\nonumber \\
 &  & \quad-\rho_{(i,j;1000)}[\rho_{(i,j;0120)}+2\rho_{(i,j;0111)}+\rho_{(i,j;0102)}]\nonumber \\
 &  & \quad-\rho_{(i,j;0100)}[\rho_{(i,j;1020)}+2\rho_{(i,j;1011)}+\rho_{(i,j;1002)}]\big\rbrace .\nonumber\end{eqnarray}
This can be further simplified by using equation (\ref{eq:pureRhoRelation}) to obtain
\begin{eqnarray}
\lambda_{1} & = & \sum_{i,j}\frac{a_{i}^{2}b_{j}^{2}}{9{\rho_{(i,j;0000)}}^{2}}[\rho_{(i,j;1100)}\rho_{(i,j;0011)}\nonumber \\
 &  & \quad+\rho_{(i,j;0000)}\rho_{(i,j;1111)}-\rho_{(i,j;1000)}\rho_{(i,j;0111)}\nonumber \\
 &  & \quad-\rho_{(i,j;0100)}\rho_{(i,j;1011)}].\label{eq:lambda1}\end{eqnarray}
Again the entanglement is completely determined by $S_v=h(\epsilon)$, where $\epsilon=\lambda_{1}$ as before. 

\subsection{Concurrence and negativity for general bipartite multi-mode pure
states}
In a similar way, we can generalize our previous expressions \cite{lin2} for the concurrence \cite{wootters98} and negativity \cite{JModOptic46.145,Zyczkowski1998} of the system after the preliminary measurement has been made.

For an $n_{1}\otimes n_{2}$ ($n_{1}\leq n_{2}$)
bipartite system, where $n_{1}$ and $n_{2}$ are Hilbert space dimension
for two subsystems respectively, the generalized concurrence of a pure quantum state
$\psi$ is defined by \cite{PRL95.040504}
\begin{equation}
\mathcal{C}^{2}(\left|\psi\right\rangle )=4\sum_{m<n}\lambda_{m}\lambda_{n},\label{eq:csquarepure}
\end{equation}
where $\sqrt{\lambda_{m}}$ ($m=1,\ldots,n_{1}$) are the eigenvalues of the reduced
density matrices $\rho^{(A)}$ and $\rho^{(B)}$. 
Additionally, the trace norm of the partial transposed density matrix
with respect to Alice's subsystem turns out to be
\begin{equation}
\|\hat{\rho}^{T_{A}}\|=(\sum_{m}\sqrt{\lambda_{m}})^{2}.\end{equation}
From this we can determine the negativity, which is defined as
\begin{equation}
\mathcal{N}(\hat{\rho})=\frac{\|\hat{\rho}^{T_{A}}\|-1}{2}.\end{equation}

As we argued earlier, the reduced density matrix in the discarding ensemble has only two non-zero eigenvalues ($\lambda_{1}$ and $(1-\lambda_{1})$) to the lowest order so we then have from equation (\ref{eq:csquarepure}):
\begin{eqnarray}
4\sum_{m<n}\lambda_{m}\lambda_{n} & = & 4\lambda_{1}+\mathrm{H.T.}\nonumber \\
 & = & \big((\sum_{m}\sqrt{\lambda_{m}})^{2}-1\big)^{2}\end{eqnarray}
where we have used $\sum_{m}\lambda_{m}=1$. Therefore, we have proved
that in the limit of small $a_i$ and $b_j$, for any multi-mode bipartite pure
state $\psi$,
\begin{equation}
\mathcal{C}(\psi)=2\mathcal{N}(\psi)=2\sqrt{\epsilon}.\label{eq:ci}
\end{equation}

Specifically, the squared concurrence is
\begin{eqnarray}\label{eq:ChighD}
\mathcal{C}^{2} & = & \sum_{ij}\left({\frac{2a_{i}b_{j}}{3|\psi|^{2}}}\right)^{2}\left|\psi\frac{\partial^{2}\psi}{\partial q_{A,i}\partial q_{B,j}}-\frac{\partial\psi}{\partial q_{A,i}}\frac{\partial\psi}{\partial q_{B,j}}\right|^{2}\\
 & \equiv & \sum_{ij}\mathcal{C}_{ij}^{2},\nonumber \end{eqnarray}
where $i$ goes over all dimensions of Alice's subsystem and $j$
of Bob's subsystem.  $\mathcal{C}_{ij}^{2}$ is the squared concurrence associated with the degrees of freedom $i$ and $j$.  Note that $\mathcal{C}_{ij}\propto a_ib_j$, consistent with the existence of a well-defined local concurrence density for two-mode systems \cite{lin2}.

Note also that the concurrence is made particularly simple by writing
\begin{equation}
\psi=e^{-S},\label{eq:expS}
\end{equation}
in which case
\begin{equation}
\mathcal{C}^{2}=\sum_{ij}\frac{4a_{i}^{2}b_{j}^{2}}{9}\left|\frac{\partial^{2}S}{\partial q_{A,i}\partial  q_{B,j}}\right|^{2}.\label{eq:squaredCforS}
\end{equation}
From this, we see that if $S$ is quadratic in the coordinates (i.e., the state is a Gaussian) the local entanglement is constant; on the other hand whenever $S$ is a linear function of the coordinates, the local entanglement is zero.

\subsection{Nodes in the wavefunction}\label{sec:nodes}
Evidently $S$ in equation~(\ref{eq:expS}) diverges near nodes of the wavefunction,  so that for a fixed $a_i$ and $b_j$ the concurrence given by equation~(\ref{eq:squaredCforS}) also diverges (like $1/|\psi|^2$ as $|\psi|\rightarrow 0$). It is important to realize that this diverging quantity refers to the entanglement in the discarding ensemble (i.e., in the sub-ensemble conditional on finding the particles in the chosen measurement region---see equation~\ref{eq:nondiscdef}), and that even in this ensemble our expression applies only in the limit of very small measurement regions.  We now show that the discarding entanglement always remains finite provided we keep within the domain of validity of our approach.

The extent of the domain of validity follows inevitably from our Taylor-series approximations for the wavefunctions (or density operators---see equation (\ref{eq:rhopsiderivs})), which are valid only close to the chosen reference point $(\mathbf{\bar{q}}_{A},\mathbf{\bar{q}}_{B})$. The requirement that the second term in this expansion be small compared with the first is
\begin{equation}\label{eq:abcondition}
\frac{\partial\psi}{\partial q_{A,i}}a_i\ll \psi(\mathbf{\bar{q}}_{A},\mathbf{\bar{q}}_{B})\quad\Rightarrow\quad a_i\ll \frac{\psi(\mathbf{\bar{q}}_{A},\mathbf{\bar{q}}_{B})}{\partial\psi/\partial q_{A,i}}
\end{equation}
and similarly for $b_j$; therefore, the domain of validity shrinks to zero near a node in $\psi$.  
Equivalently, if this condition is not satisfied it leads to the breakdown of the isomorphism of each mode to one qubit described in \cite{lin2}.

One way to understand the behavior of the entanglement near points where the wavefunction vanishes is to satisfy equation~(\ref{eq:abcondition}) by writing the maximum valid region size as
\begin{equation}
a_i^\mathrm{MAX} = \sigma\frac{\psi(\mathbf{\bar{q}}_{A},\mathbf{\bar{q}}_{B})}{\partial\psi/\partial q_{A,i}},
\end{equation}
where $\sigma \ll 1$ is a small parameter, and similarly for $b_j^\mathrm{MAX}$.   (We assume here that the derivatives are not also zero near the nodes.) We further define three quantities $k_i$, $k_j$, and $k_{ij}$ by
\begin{eqnarray}
\frac{\partial^{2}\psi}{\partial q_{A,i}\partial q_{B,j}} &=& k_{ij} \psi; \nonumber \\
\frac{\partial\psi}{\partial q_{A,i}} &=& k_i \psi; \nonumber \\
\frac{\partial\psi}{\partial q_{B,j}} &=& k_j \psi,
\end{eqnarray}
so $a_i^\mathrm{MAX}k_i=b_j^\mathrm{MAX}k_j=\sigma$.
From equation~(\ref{eq:ChighD}), if we choose $a_i=a_i^\mathrm{MAX}$, $b_j=b_j^\mathrm{MAX}$ near a node where $k_ik_j\gg k_{ij}$, the expression for $\epsilon$ reduces to
\begin{equation}\label{eq:epsilonmax}
\epsilon_\mathrm{MAX} = \sum_{ij}\frac{\sigma^4}{9}.
\end{equation}
Therefore $\epsilon$ (and hence also the localized concurrence and entanglement) is cut off near the node at a finite value that depends on the choice of $\sigma$.

%There are therefore two ways to understand the behavior of the entanglement near points where the wavefunction vanishes.  One is to remember that our measurement region always has to satisfy (\ref{eq:abcondition}) and therefore its maximum size must shrink as we approach the node, preventing any divergence in the local entanglement.  Another is to recognize that we are anyway unlikely to find the system in regions where the wavefunction is tending to zero; if we weight the local entanglement by the probability of finding the system in the region, we are in effect working with the nondiscarding entanglement $\mathcal{E}_{ND}$ (see equation~(\ref{eq:nondiscdef})).  We may also weight other quantities, such as the concurrence or negativity, similarly; these are useful because their divergence in the discarding ensemble is precisely cancelled as the probability of finding the particle tends to zero.

\subsection{Transformation of coordinates}

We now discuss the behavior of our expressions for the local entanglement under various coordinate transformations.

\subsubsection{Invariance under local transformations}
We would expect that the definitions of our local entanglement would remain unchanged if we made a local redefinition of our coordinate axes (possibly accompanied by changes in the measurement region).  To see that this is the case, consider the following transformation of Alice's coordinates:
\begin{equation}\label{eq:definetransform}
\frac{Q_{i}}{A_i}=\sum_{j}O_{ij}\frac{q_{j}}{a_j}\end{equation}
where $O$ is an orthogonal matrix ($OO^{T}=\mathbf{1}$) and the sum goes only over the other coordinates of Alice's particle.   $A_i$ is to determine the length of the measurement region for new variable $Q_i$.  Note that if $a_j=A_i=a\;\forall i,j$ (i.e. both measurement volumes are hypercubes with the same dimensions) then (\ref{eq:definetransform}) reduces to a simple orthogonal transformation of Alice's coordinates.

Now
\begin{equation}
a_i\frac{\partial}{\partial q_{i}}=\sum_{j}A_j\frac{\partial Q_{j}}{\partial q_{i}}\frac{\partial}{\partial Q_{j}}=\sum_{j}O_{ij}A_j\frac{\partial}{\partial Q_{j}}.\end{equation}
We then have

\begin{eqnarray}
\sum_{i}a_i^2\frac{\partial^{2}\rho}{\partial q_{i}\partial{q'}_{i}} & =& \sum_{ijk}O_{ij}O_{ik}A_jA_k\frac{\partial^{2}\rho}{\partial Q_{j}\partial{Q'}_{k}} \nonumber \\
& =& \sum_{j}A_j^2\frac{\partial^{2}\rho}{\partial Q_{j}\partial{Q'}_{j}}\end{eqnarray}
and similarly

\begin{equation}
\sum_{i}a_i^2\frac{\partial\rho}{\partial q_{i}}\frac{\partial\rho}{\partial{q'}_{i}}=\sum_{i}A_i^2\frac{\partial\rho}{\partial Q_{i}}\frac{\partial\rho}{\partial{Q'}_{i}}.\end{equation}
Therefore, equation~(\ref{eq:multiDlambdas})  is invariant under the generalized orthogonal transformation
(\ref{eq:definetransform}). It follows that equation~(\ref{eq:lambda1}), and hence the local entanglement, are also  invariant under these local transformations. 

\subsubsection{Non-local transformations}\label{sec:nonlocal-transformation}
We now consider some transformations which mix Alice's and Bob's coordinates---specifically, those that  make the system separable.  That is to say we look for a new set of coordinates
\begin{equation}\label{eq:transformdef}
X_k=\sum_{i} T_{ik} x_{i}
\end{equation}
such that the wavefunction factorizes as
\begin{equation}
\psi=\prod_{k} \psi_{k}(X_k).
\end{equation}
Note that the sum over $i$ in (\ref{eq:transformdef}) runs over all coordinates of the system (both Alice's and Bob's).  In this situation it does not make sense to consider any accompanying change in the shape or size of the measurement region, which we continue to define in terms of the original coordinates and to describe by $\{a_i\}$ and $\{b_j\}$.

Therefore,
\begin{eqnarray}
\frac{\partial^{2}\psi}{\partial x_{i}\partial{x}_{j}} & =& \sum_{kk'}T_{ik}T_{jk'}
\frac{\partial^{2}\psi}{\partial X_{k}\partial{X}_{k'}} \\
& =& \sum_{k}T_{ik}T_{jk}\frac{\psi}{\psi_k}\frac{\partial^{2}\psi_k}{\partial X_{k}^2}\nonumber  \\
&   &\quad +\sum_{k\ne k'}T_{ik}T_{jk'}\frac{\psi}{\psi_k \psi_k'}
\frac{\partial\psi_k}{\partial X_{k}}\frac{\partial\psi_{k'}}{\partial{X}_{k'}}\nonumber  \end{eqnarray}
and similarly
\begin{equation}\label{eq:productoffirstderiv}
\frac{\partial\psi}{\partial x_{i}}\frac{\partial\psi}{\partial{x}_{j}}=\sum_{kk'}T_{ik}T_{jk'}\frac{\psi^2}{\psi_k \psi_k'}\frac{\partial\psi_k}{\partial X_{k}}\frac{\partial\psi_{k'}}{\partial{X}_{k'}}.
\end{equation} 
It follows from equation~(\ref{eq:ChighD}) that
\begin{equation}\label{eq:epsilonfortrasformedx}
\epsilon= \sum_{ij}\frac{(a_{i}b_{j})^{2}}{9}\left|\sum_{k}T_{ik}T_{jk}\frac{\psi}{\psi_k}\left[\frac{\partial^{2}\psi_k}{\partial X_{k}^2}-\frac{1}{\psi_k}
\left(\frac{\partial\psi_k}{\partial X_{k}}\right)^2\right]\right|^{2},
 \end{equation}
 where the second term inside the modulus signs comes from the part of (\ref{eq:productoffirstderiv}) having $k=k'$.  In terms of the logarithms of the separable wavefunctions $S_k=-\log[\psi_{k}(X_k)])$,
we have
\begin{equation}\label{eq:nonlocalepsilon}
\epsilon=\sum_{ij}\frac{(a_{i}b_{j})^{2}}{9}\left|\sum_{k}T_{ik}T_{jk} \frac{\partial^2 S_k}{\partial X_{k}^2}  \right|^{2}.
\end{equation}

One important special case of this result is the transformation to normal coordinates in a harmonic system: if the potential can be quadratically expanded about an energy minimum, the transformation to normal coordinates takes the form of equation~(\ref{eq:transformdef}) with 
\begin{equation}
T_{ik}=\sqrt{m_i}O_{ik},
\end{equation}
where $O$ is an orthogonal matrix.

\subsubsection{Relative coordinates}\label{sec:relative-coordinates}
A closely related example is the transformation to center-of-mass and relative coordinates.  (Here we assume that the particles live in the same physical space, and hence that the dimensions $N_A$ and $N_B$ are equal.) If Alice's particle and Bob's particle have masses $m_{A}$ and $m_{B}$ respectively, we define $r_{i}\equiv q_{i}^{A}-q_{i}^{B}$ and $R_{i}\equiv(\mu/m_{B})q_{i}^{A}+(\mu/m_{A})q_{i}^{B}$
where $\mu\equiv m_{A}m_{B}/(m_{A}+m_{B})$ is the reduced mass and $i$ goes over all dimensions of the system ($\{ x,y,z\}$ in three-dimensional system, for example).  
\begin{eqnarray}
\epsilon & = & \sum_{ij}\left({\frac{a_{i}b_{j}}{3|\psi|^{2}}}\right)^{2}\left|-(\frac{\mu}{m_{B}}\frac{\partial\psi}{\partial R_{i}}+\frac{\partial\psi}{\partial r_{i}})(\frac{\mu}{m_{A}}\frac{\partial\psi}{\partial R_{j}}-\frac{\partial\psi}{\partial r_{j}})\right.\nonumber \\
 &  & _{\quad}\left.+\psi(\frac{\mu}{m_{B}}\frac{\partial}{\partial R_{i}}+\frac{\partial}{\partial r_{i}})(\frac{\mu}{m_{A}}\frac{\partial}{\partial R_{j}}-\frac{\partial}{\partial r_{j}})\psi\right|^{2},\label{eq:withCOMepsilon}\end{eqnarray}
where $i$ and $j$ run over all the dimensions of the system.

In many cases, including most importantly the case where there is no external potential, the wave function $\psi(\mathbf{R},\mathbf{r})$ can be decoupled into a center-of-mass part $\chi(\mathbf{R})$ and a relative-motion part $\varphi(\mathbf{r})$:
\begin{equation}
\psi(\mathbf{R},\mathbf{r})=\chi(\mathbf{R})\varphi(\mathbf{r}).
\end{equation}
If we write 
\begin{equation}
\varphi(\mathbf{r})=e^{-S_\varphi(\mathbf{r})},\qquad \chi(\mathbf{R})=e^{-S_\chi(\mathbf{R})}\label{eq:defineOmega}
\end{equation}
then the entanglement takes the particularly simple form
\begin{equation}\label{eq:epsilonTotal2}
\epsilon=\sum_{ij}\frac{a_{i}^{2}b_{j}^{2}}{9}\left|\frac{\partial^{2}S_\varphi(\mathbf{r})}{\partial r_{i}\partial r_{j}}+\frac{\mu^2}{m_Am_B}\frac{\partial^2S_\chi}{\partial R_i\partial R_j}\right|^{2}.\end{equation}
For example, if $\chi(\mathbf{R})$ is a free-particle plane wave $\chi(\mathbf{R})=e^{i\mathbf{k}_{0}\mathbf{R}}$, its contribution
to the entanglement $\mathcal{E}_{D}$ is zero; if $\chi(\mathbf{R})$ is a 
Gaussian wave packet with wave number $\mathbf{k}_{0}$ and real-space width $R_0$:
\begin{equation}\label{eq:gaussianpacket}
\psi(\mathbf{R},\mathbf{r})=(\frac{2}{\pi R_{0}^{2}})^{1/4}e^{-\mathbf{R}^{2}/R_{0}^{2}}e^{i\mathbf{k}_{0}\mathbf{R}}\varphi(\mathbf{r}),\end{equation}
the expression for $\epsilon$ becomes
\begin{equation}
\epsilon=\sum_{ij}\frac{a_{i}^{2}b_{j}^{2}}{9}\left|\frac{\partial^{2}S_\varphi(\mathbf{r})}{\partial r_{i}\partial r_{j}}-\frac{2\mu^{2}}{m_{A}m_{B}R_{0}^{2}}\delta_{ij}\right|^{2}.\label{eq:epsilonTotalgaussian}
\end{equation}

\section{Examples}\label{sec:examples}
In this section we apply our method to some easily soluble examples: first to wavefunctions that (while remaining pure states) are semiclassical in the sense that the potential varies slowly on the scale of the de Broglie wavelength, so WKB methods are applicable, then to energy eigenstates of harmonically-interacting particles in arbitrary dimensionality, and finally to bound states of an electron and proton (i.e., to the hydrogen atom).

\subsection{The semiclassical case: one-dimensional WKB wavefunctions}
% May want to change to 1D
Consider two particles moving in one dimension with an interaction potential $V(r)$ that depends only on the relative coordinate.  Neglecting center-of-mass contributions, the entanglement can then be calculated from the relative wavefunction $\varphi(r)$ using equation~(\ref{eq:epsilonTotal2}).
If $V(r)$ is a slowing varying function of $r$, we can use the
WKB method to find $\varphi(r)$.

{\begin{figure} 
\includegraphics[width=0.8\columnwidth,keepaspectratio]{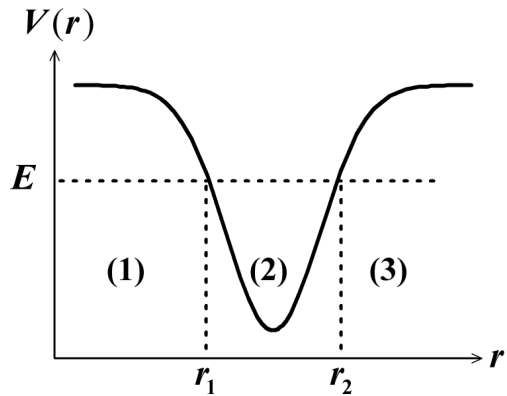} 
\caption{\label{fig:regionlabels}
Diagram of a potential well illustrating the different regions discussed in the text.}
\end{figure} 

We consider an interaction with a single potential well (shown schematically in Figure~\ref{fig:regionlabels}), so the system moving in a bound state with energy $E$ has just two classical turning points.  For the classically allowed region with $E>V$ (region 2 of Figure~\ref{fig:regionlabels}), the classical momentum at $r$ is $p(r)=\sqrt{2m(E-V(r))}$ and the corresponding wavefunction can be expressed as
\begin{equation}
\varphi_{2}^{WKB}(r)= \frac{2A}{\sqrt{p(r)}}\sin\left[\frac{1}{\hbar}\int_{r}^{r_{2}}p(r')\,\mathrm{d}r'+\frac{\pi}{4}\right],\quad r_{1}<r<r_{2}
\end{equation} 
so that the local concurrence is
\begin{eqnarray}\label{eq:concurrencer2}
\mathcal{C}_{2} & = &\biggl| \frac{ab}{3\hbar^{2}p(r)^2}\biggl\{2\csc^2\left[\frac{1}{\hbar}\int_{r}^{r_{2}}p(r')\mathrm{d}r'+\frac{\pi}{4}\right]p(r)^{4}\nonumber \\
 &  & \quad+\hbar^{2}p(r)\frac{\partial^{2}p(r)}{\partial r^{2}}\biggr\}-\hbar^{2}\left(\frac{\partial p(r)}{\partial r}\right)^{2}\\
 &  & \quad+2\hbar\cot\left[\frac{1}{\hbar}\int_{r}^{r_{2}}p(r')\mathrm{d}r'+\frac{\pi}{4}\right]p(r)^{2}\frac{\partial p(r)}{\partial r}\biggr|.\nonumber 
 \end{eqnarray}
The oscillatory structure of the wavefunction, arising from the interference between right- and left-moving travelling waves, produces nodes at which the entanglement in the discarding ensemble for fixed $a$ and $b$ diverges (but remains finite provided we remain within the domain of validity of (\ref{eq:concurrencer2})---see \S\ref{sec:nodes}).  

Note also that the entanglement contribution from the first term in (\ref{eq:concurrencer2}) is non-zero even where $V(r)$ (and hence $p(r)$) is constant.

For $E<V$ (region 1 and region 3 of Figure~\ref{fig:regionlabels}), we express the wavefunction in terms of the local momentum on the inverted potential surface $p(r)=\sqrt{2m(V(r)-E)}$.  The wavefunctions are respectively
\begin{eqnarray}
\varphi_{1}^{WKB}(r) & = & \frac{(-1)^{n}A}{\sqrt{\left|p(r)\right|}}\exp\left[-\frac{1}{\hbar}\int_{r}^{r_{1}}|p(r')|\mathrm{d}r'\right],\quad r<r_{1};\nonumber \\
 &  &\\
\varphi_{3}^{WKB}(r) & = & \frac{A}{\sqrt{\left|p(r)\right|}}\exp\left[-\frac{1}{\hbar}\int_{r_{2}}^{r}|p(r')|\mathrm{d}r'\right],\quad r>r_{2},\nonumber \\
 &  &
\end{eqnarray}
where $n$ is the number of nodes in Region 2.  Correspondingly, the concurrences are
\begin{eqnarray}
\mathcal{C}_{1} & = & \biggl|\frac{-ab}{3\hbar\left|p(r)\right|^{2}}\biggl[2\left|p(r)\right|^{2}\frac{\partial\left|p(r)\right|}{\partial r}\nonumber \\
&  & \quad+\hbar\left(\frac{\partial\left|p(r)\right|}{\partial r}\right)^{2}-\hbar\left|p(r)\right|\frac{\partial^{2}\left|p(r)\right|}{\partial r^{2}}\biggr]\biggr|; \\
\mathcal{C}_{3} & = & \biggl|\frac{ab}{3\hbar\left|p(r)\right|^{2}}\biggl[2\left|p(r)\right|^{2}\frac{\partial\left|p(r)\right|}{\partial r}\nonumber \\
&  & \quad-\hbar\left(\frac{\partial\left|p(r)\right|}{\partial r}\right)^{2}+\hbar\left|p(r)\right|\frac{\partial^{2}\left|p(r)\right|}{\partial r^{2}}\biggr]\biggr|.\end{eqnarray}
Note that in this case (by contrast to the behavior in region 2) if there is no \emph{force}, $p(r)$ is constant, and hence there is no entanglement.  It is interesting that the boundaries between these different behaviors of the entanglement correspond to the classical turning points.

\subsection{Multi-dimensional harmonic oscillators}
Consider first a system of two one-dimensional harmonic oscillators of masses $m_A$ and $m_B$, having identical frequencies $\omega$, and coupled  by a spring constant $K$; the Hamiltonian is
\begin{equation}
\hat{H}=\hat{H}_{A}+\hat{H}_{B}+\frac{1}{2}K(\hat{X}_{A}+\hat{X}_{B})^2.\label{eq:Hfor2HO}
\end{equation}
Transforming to center-of-mass and relative coordinates, the eigenstates are simply
\begin{eqnarray}
\psi_{n_R,n_r}(R,r) & = & \psi_{n_R}(R) \psi_{n_r}(r)  \nonumber \\
                       & = &  \frac{1}{\sqrt{\sqrt{\pi}2^{n_R}2^{n_r}n_{R}!n_{r}!R_{0}r_{0}}}e^{-R^{2}/2R^2_{0}}\nonumber \\
                       &   & \quad e^{-r^{2}/2r^2_{0}}H_{n_R}(\frac{R}{R_0}) H_{n_r}(\frac{r}{r_0}),
\end{eqnarray}
where $n_R$ and $n_r$ label the excitations of each coordinate, $R_0=\sqrt{\hbar/(M\omega)}$, $r_0=\sqrt{\hbar/(\mu\sqrt{\omega^2+K/\mu})}$, and $H_n(x)$ is the Hermite polynomial.

If Alice and Bob each possess an oscillator, the entanglement between their subsystems  given by $h(\epsilon)$ can be determined from equation (\ref{eq:withCOMepsilon}); for example, for the ground state:
\begin{eqnarray}
\epsilon  & = & \frac{a^2b^2(m_A m_B r_0^2-M^2 R_0^2)^2}{9M^4 r_0^4 R_0^4}\nonumber \\
                   & = & \frac{a^2b^2}{9M^2 \hbar^2}(m_A m_B\omega-M\mu \sqrt{\frac{K}{\mu}+\omega^2})^2,
\end{eqnarray}
where $M=m_A+m_B$. Note that the ground state is Gaussian, so $\epsilon$ is constant, as expected.

\begin{figure}

\includegraphics[width=0.33\linewidth,keepaspectratio]{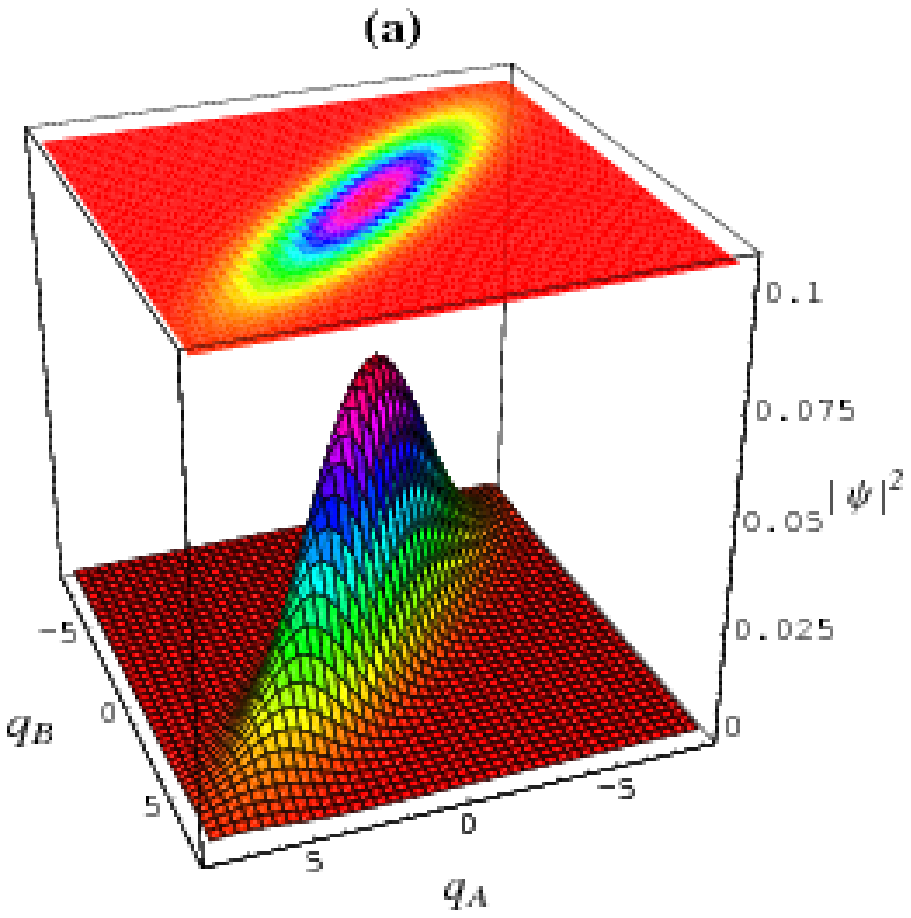}\includegraphics[width=0.33\linewidth,keepaspectratio]{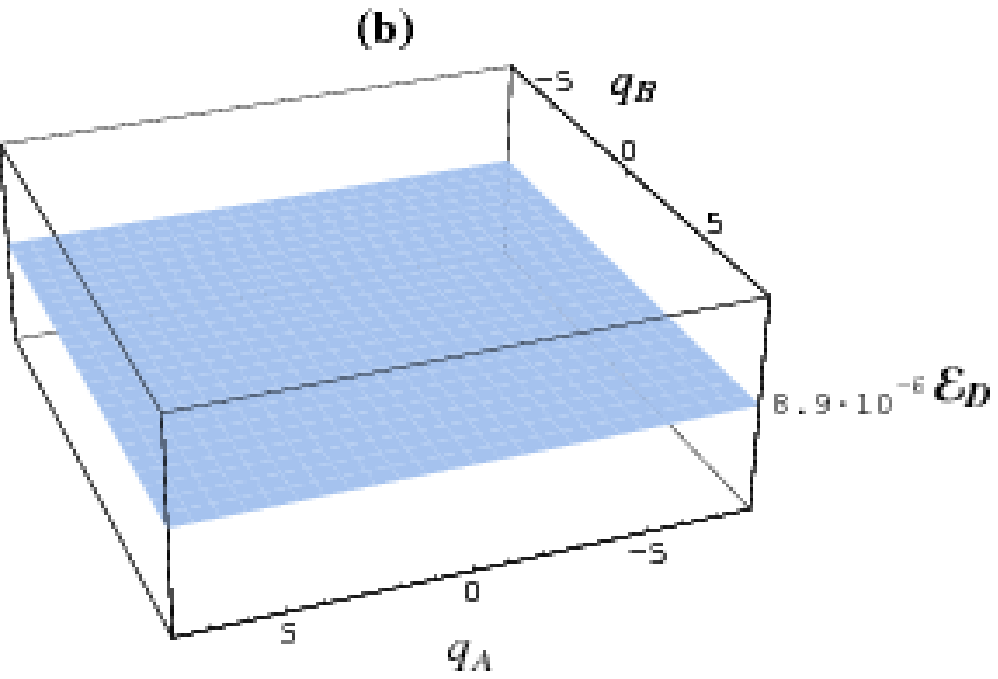}\includegraphics[width=0.33\linewidth,keepaspectratio]{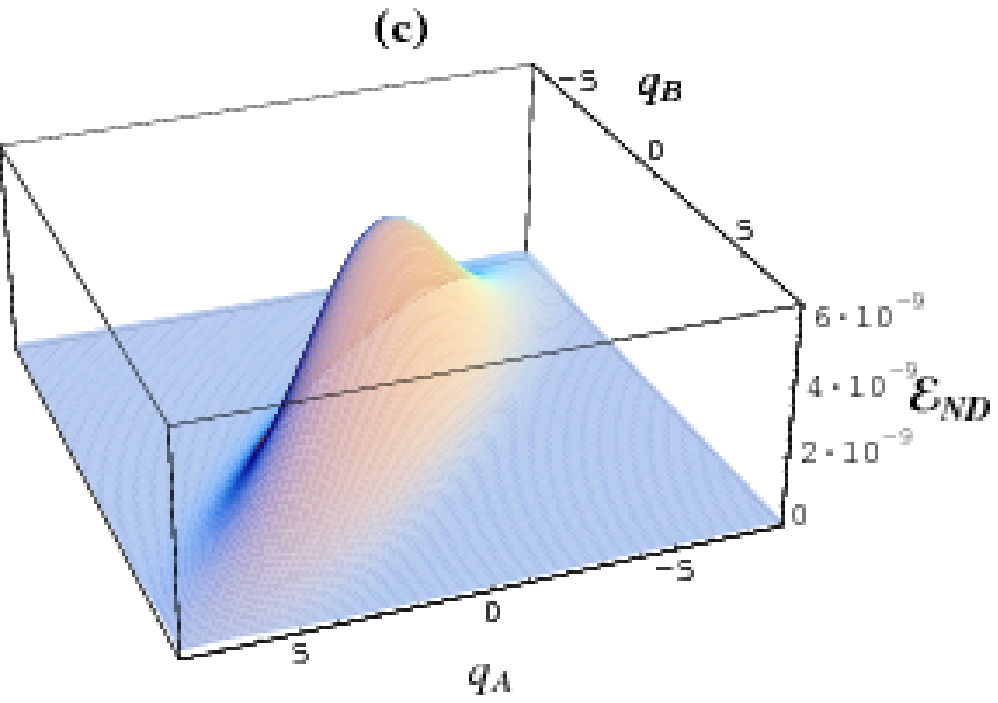}
\begin{centering}(A) $\psi_{0,0}(R,r)$\par\end{centering}

\includegraphics[width=0.33\linewidth,keepaspectratio]{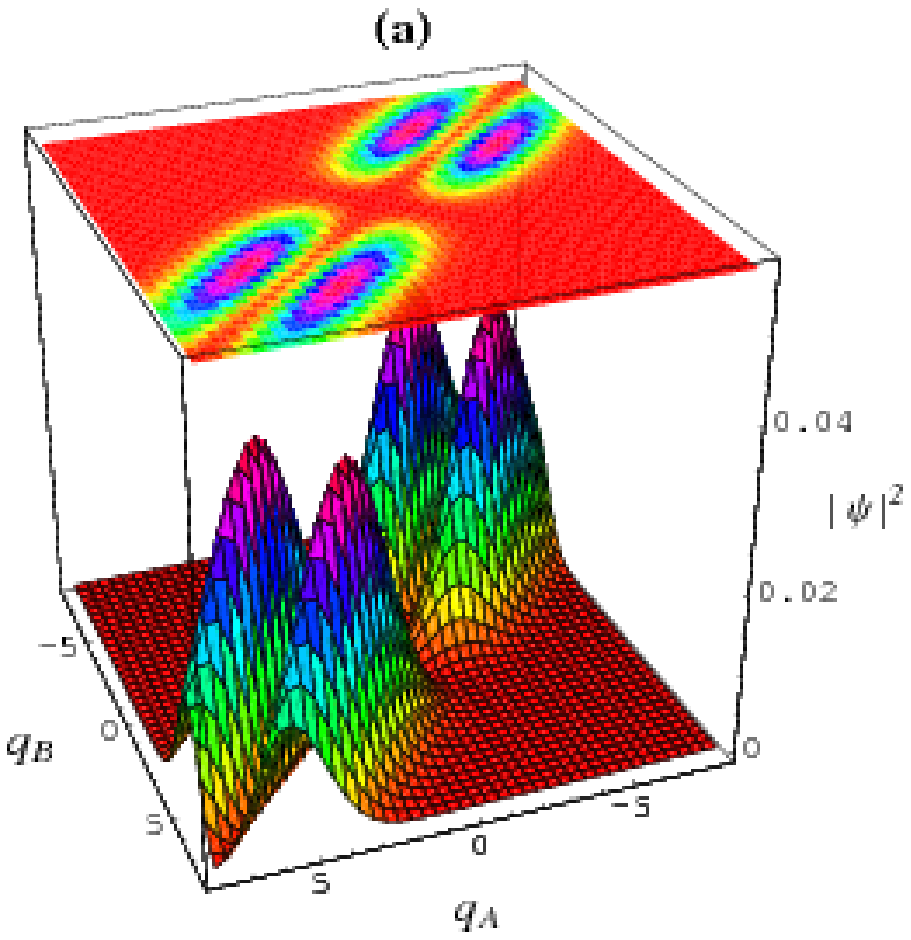}\includegraphics[width=0.33\linewidth,keepaspectratio]{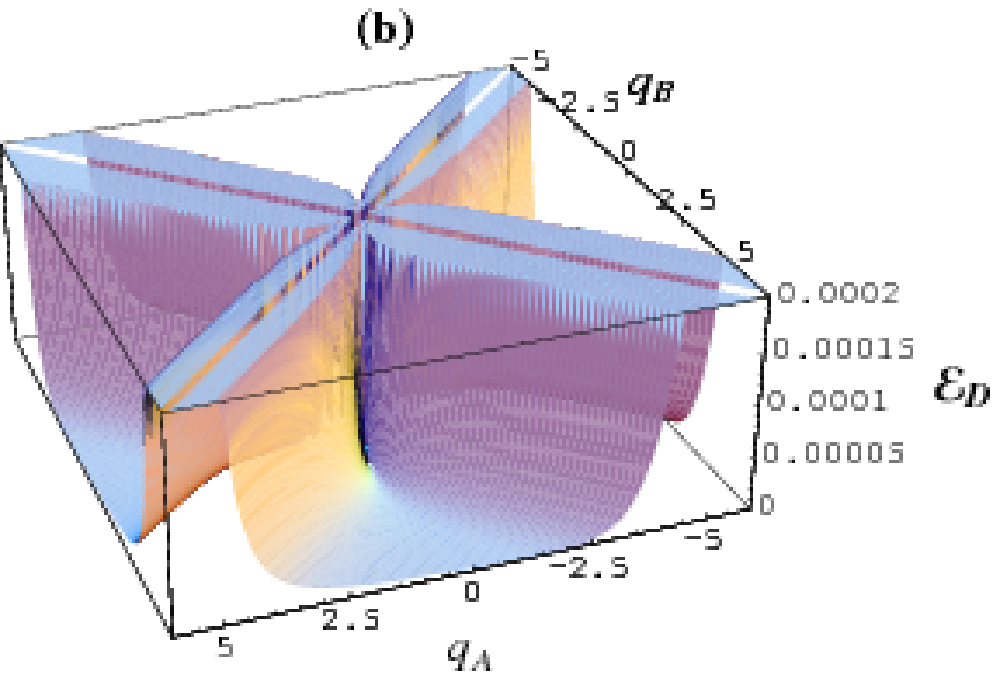}\includegraphics[width=0.33\linewidth,keepaspectratio]{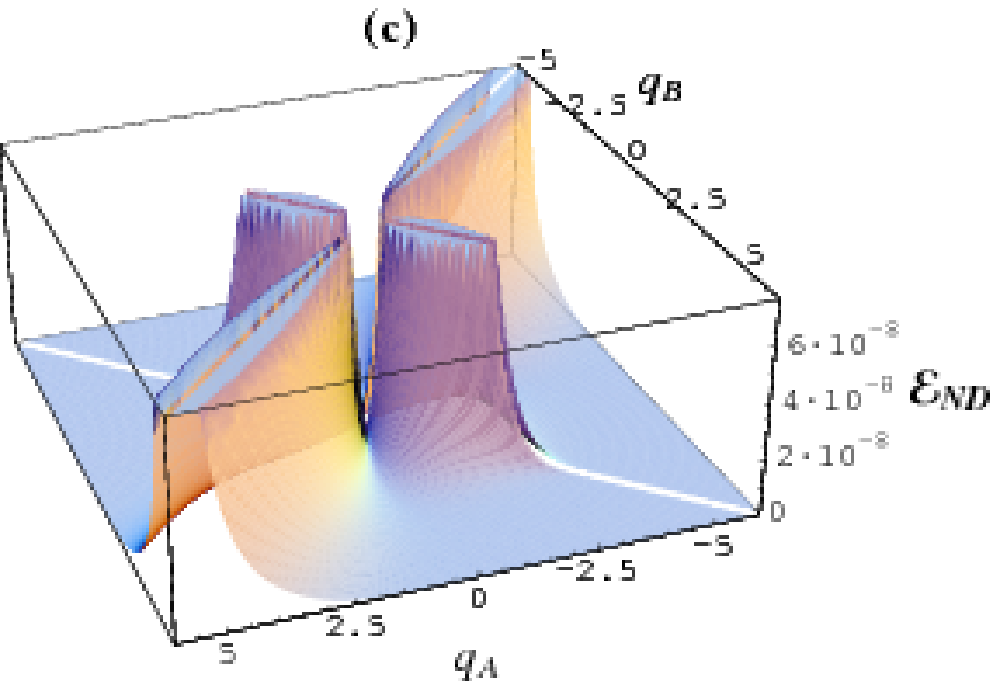}
\begin{centering}(B) $\psi_{1,1}(R,r)$\par\end{centering}

\includegraphics[width=0.33\linewidth,keepaspectratio]{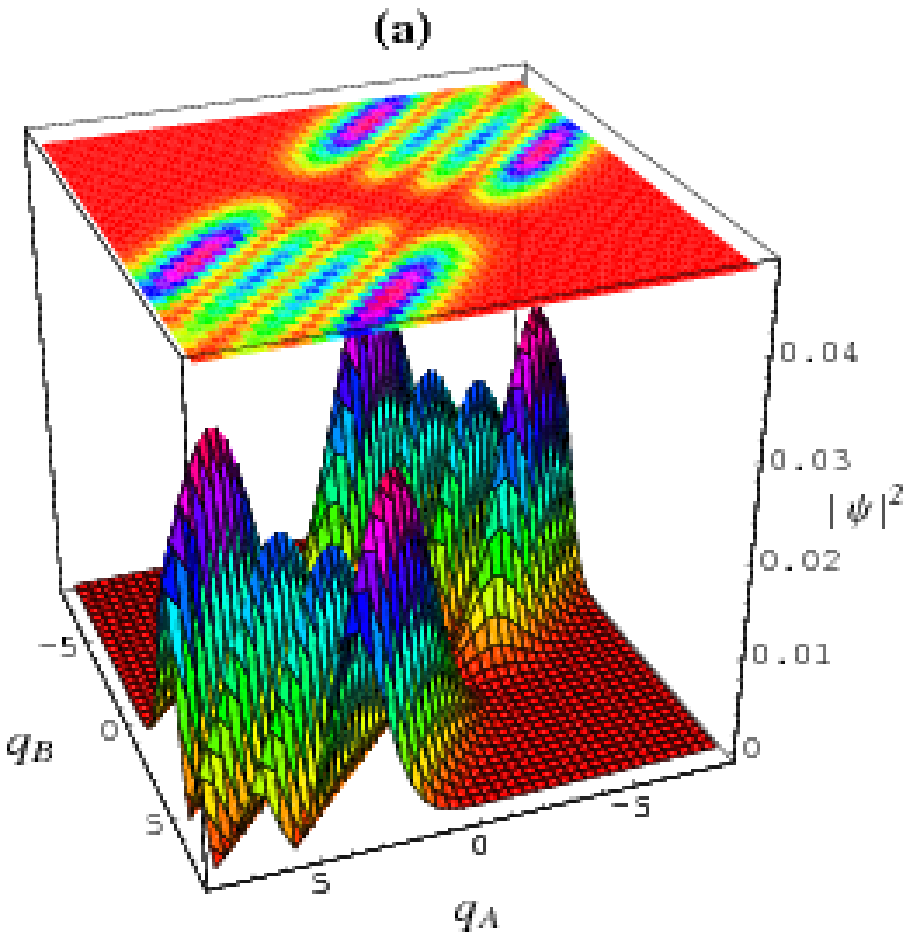}\includegraphics[width=0.33\linewidth,keepaspectratio]{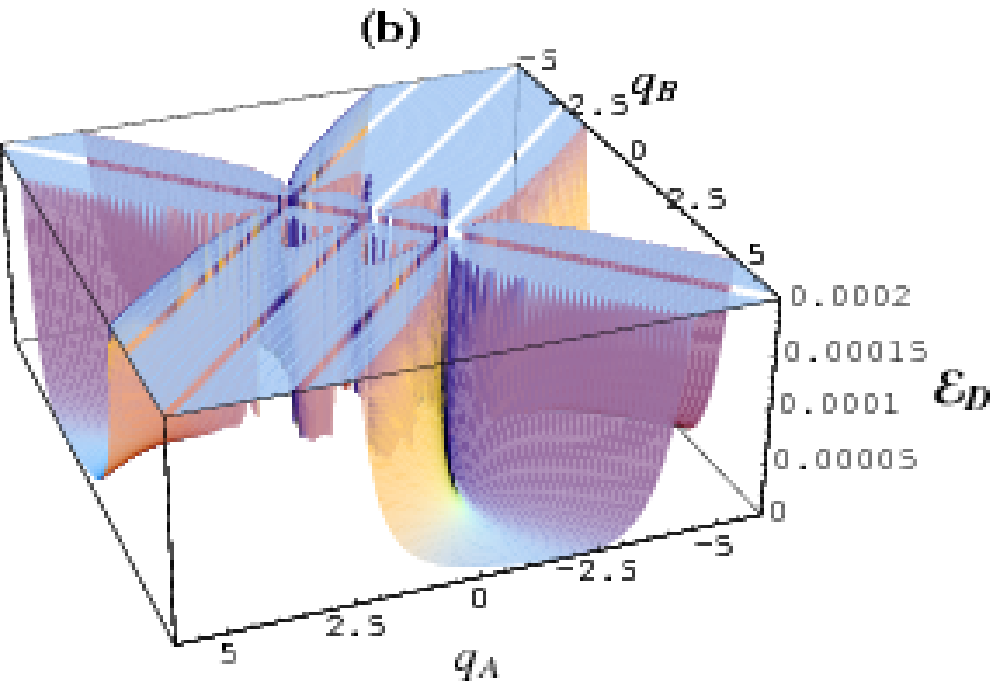}\includegraphics[width=0.33\linewidth,keepaspectratio]{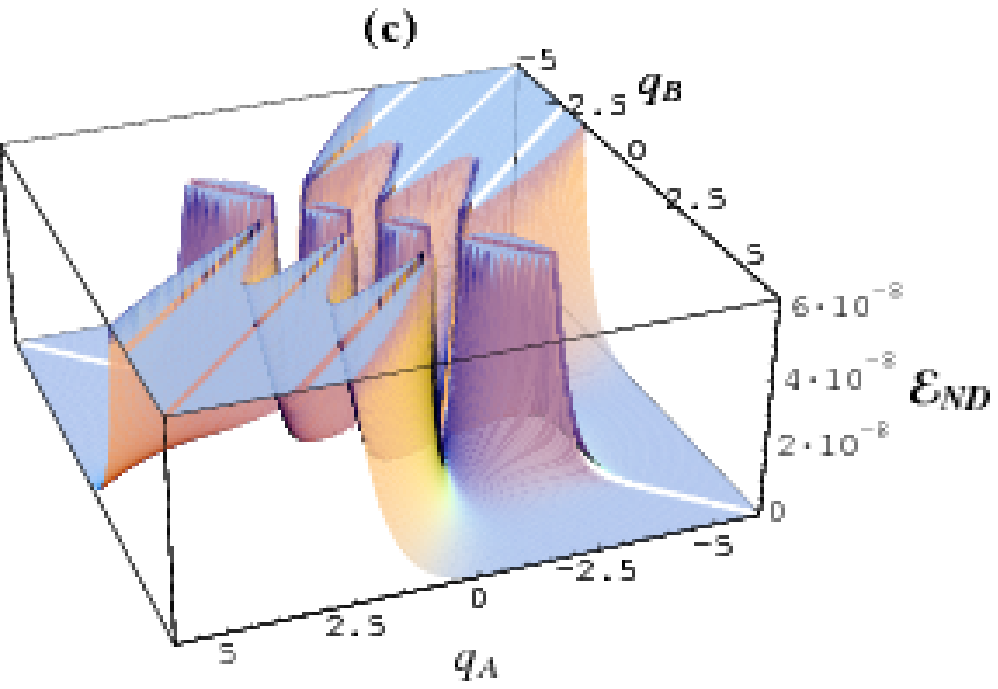}
\begin{centering}(C) $\psi_{1,3}(R,r)$\par\end{centering}

\caption{Probability density (left plot), local entanglement $\mathcal{E}_{D}$ in the discarding ensemble (center plot) and  local entanglement $\mathcal{E}_{ND}$ in the nondiscarding ensemble (right plot) for three pure states of the two-oscillator system: (A) $n_R=0$, $n_r=0$; (B) $n_R=1$, $n_r=1$; (C) $n_R=1$, $n_r=3$.   The characteristic lengths of the problem are $r_0=2$ and $R_0=4$ in all plots, and all plots are for $a=b=0.1$. The cut-off points for plots of $\mathcal{E}_{D}$ and $\mathcal{E}_{ND}$ are determined  from $\epsilon_\mathrm{MAX}$ in equation~(\ref{eq:epsilonmax})  with $\sigma=0.1$; specifically, $\mathcal{E}_{ND}^\mathrm{MAX}=p_{ab}^\mathrm{MAX} h(\epsilon_\mathrm{MAX})$, where $h$ is defined in equation~(\ref{eq:vonNeuman}).
\label{fig:HOplots}}
\end{figure}

In Fig. \ref{fig:HOplots}, we plot the probability distributions and entanglement $\mathcal{E}$ (in the discarding ensemble---center column, and nondiscarding ensemble---right column) for the ground state and some excited states.  Note that the ground state (a) is a Gaussian state so the discarding entanglement is constant and the left and right plots are proportional to one another; this is no longer true for the other (non-Gaussian) states,  for which there are also nodes in the wavefunctions.  We therefore show the entanglement in both ensembles cut off at the maximum value determined by equation~(\ref{eq:epsilonmax}).

For general multi-dimensional oscillators, the wavefunction becomes a product over the normal modes $X_k$ of one-dimensional harmonic oscillator wavefunctions.  The entanglement is determined by these normal-mode wavefunctions through equation~(\ref{eq:nonlocalepsilon}).  (Note that in the one-dimensional example considered above, the normal coordinates are the same as the relative and center-of-mass coordinates.)
 
\subsection{The hydrogen atom}
We next consider the entanglement between the electron (`Alice's particle') and the proton (`Bob's particle') in a hydrogen atom.
For simplicity, the sizes of the measured regions are
assumed to be the same for all dimensions \{$x$,
$y$, $z$\}, i.e. $a_{i}=a$ and $b_{i}=b$.
First, consider the case where there is no center-of-mass motion.
Instead of directly applying equation (\ref{eq:epsilonTotal2}),
we transform the coordinates and the equation into to spherical coordinates:
\begin{eqnarray}
\frac{\partial}{\partial r_{x}} & = & \sin\theta\cos\phi\frac{\partial}{\partial r}+\frac{\cos\theta\cos\phi}{r}\frac{\partial}{\partial\theta}-\frac{\csc\theta\sin\phi}{r}\frac{\partial}{\partial\phi}\nonumber \\
\frac{\partial}{\partial r_{y}} & = & \sin\theta\sin\phi\frac{\partial}{\partial r}+\frac{\cos\theta\sin\phi}{r}\frac{\partial}{\partial\theta}+\frac{\csc\theta\cos\phi}{r}\frac{\partial}{\partial\phi}\nonumber \\
\frac{\partial}{\partial r_{z}} & = & \cos\theta\frac{\partial}{\partial r}-\frac{\sin\theta}{r}\frac{\partial}{\partial\theta}.\label{eq:SphericalTransform}\end{eqnarray}

The ground state is 
\begin{equation}
\varphi_{100}(r,\theta,\phi)=(\frac{1}{\pi a_{0}^{3}})^{1/2}e^{-r/a_{0}},\end{equation}
where $a_{0}$ is the Bohr radius. In this case,
\begin{equation}
\epsilon=2(\frac{ab}{3a_{0}r})^{2}.\end{equation}
Interestingly, this expression indicates that the entanglement $\mathcal{E}_{D}$
for the ground state of a hydrogen atom falls off with distance in exactly the same way as the
electrostatic force between the electron and the nucleus. 

If we include a center-of-mass part to the wave function with a Gaussian form as in equation~(\ref{eq:gaussianpacket}), we obtain
\begin{eqnarray}
\epsilon & = & \frac{2a^{2}b^{2}}{9R_{0}^{4}a_{0}^{2}(m_{A}+m_{B})^{4}r^{2}}\Big(R_{0}^{4}(m_{A}+m_{B})^{4}\\
 &  & -4R_{0}^{2}a_{0}m_{A}m_{B}(m_{A}+m_{B})^{2}r+6a_{0}^{2}m_{A}^{2}m_{B}^{2}r^{2}\Big).\nonumber 
\end{eqnarray}
The first term is the component noted previously, decaying in the same way as the atom's internal electrostatic force; in addition there are two new contributions from the localization of the free-particle wave function.  Of these the third term corresponds to the spatially constant entanglement of the gaussian center-of-mass state.

Excited states of the atom can also be analyzed, by substituting
the most general form of the relative wave function $\varphi_{nlm}(r,\theta,\phi)$
of a hydrogen atom  into equation (\ref{eq:epsilonTotal2})
after it has been transformed to spherical coordinates. The excited states have nodes in the wavefunction, which have to be treated  as discussed earlier.
We show the corresponding probability distribution, and entanglement $\mathcal{E}$ (in the discarding and nondiscarding ensembles) in Fig. \ref{fig:Hatomplots}.

\begin{figure}
\includegraphics[width=0.33\linewidth,keepaspectratio]{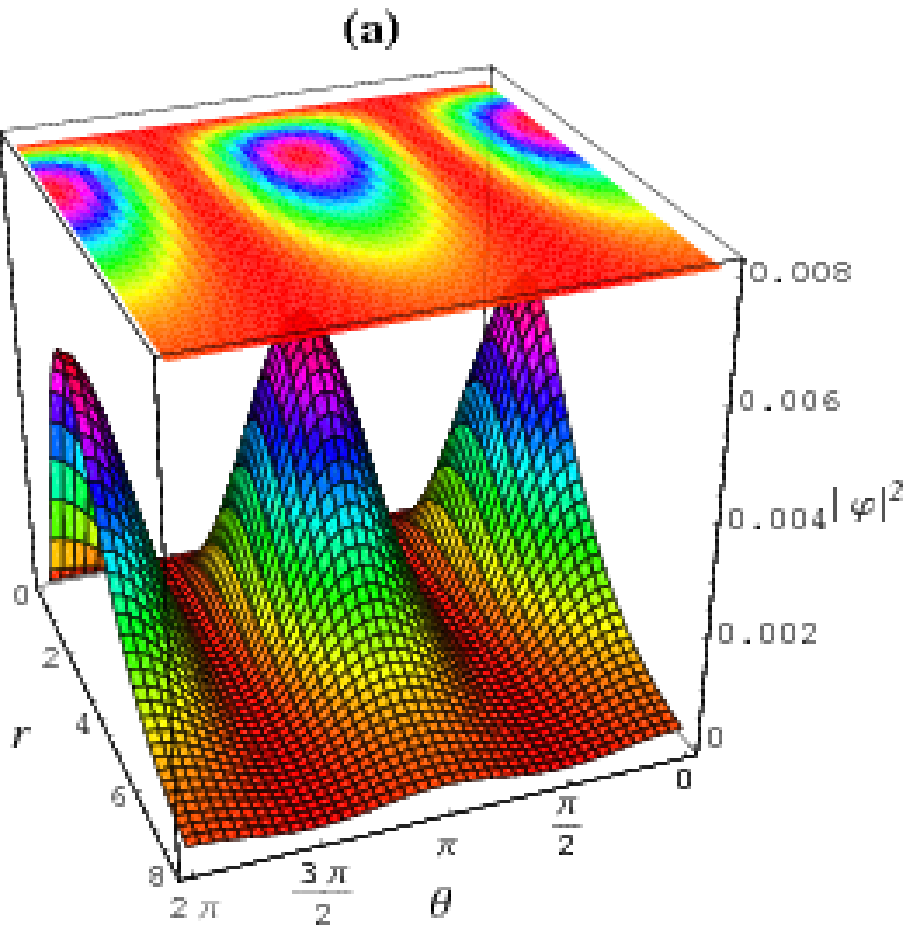}\includegraphics[width=0.33\linewidth,keepaspectratio]{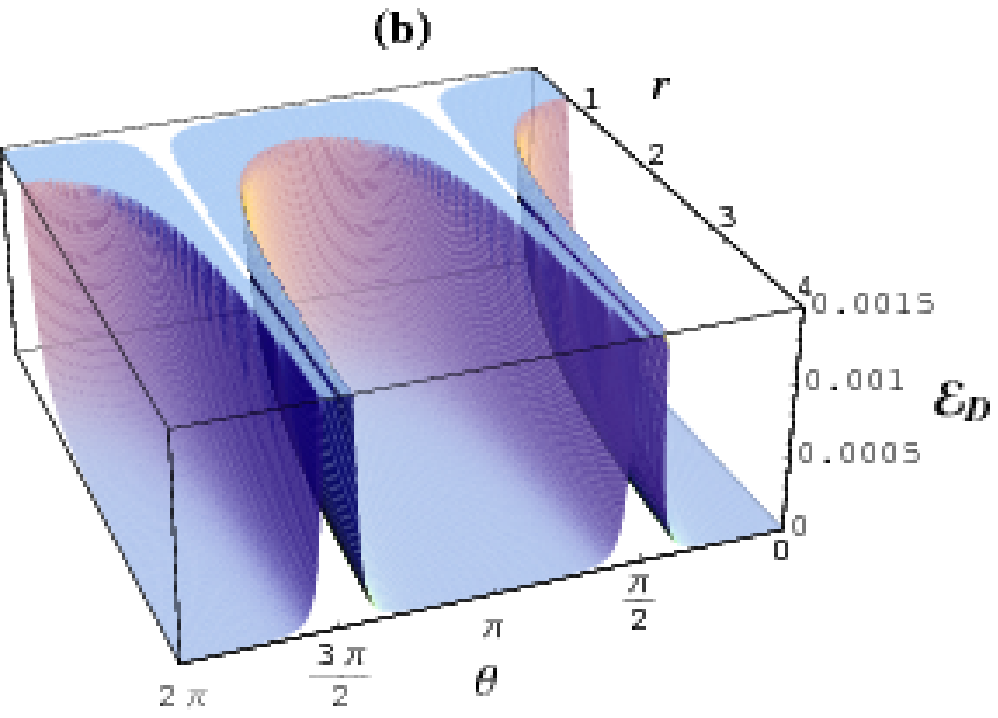}\includegraphics[width=0.33\linewidth,keepaspectratio]{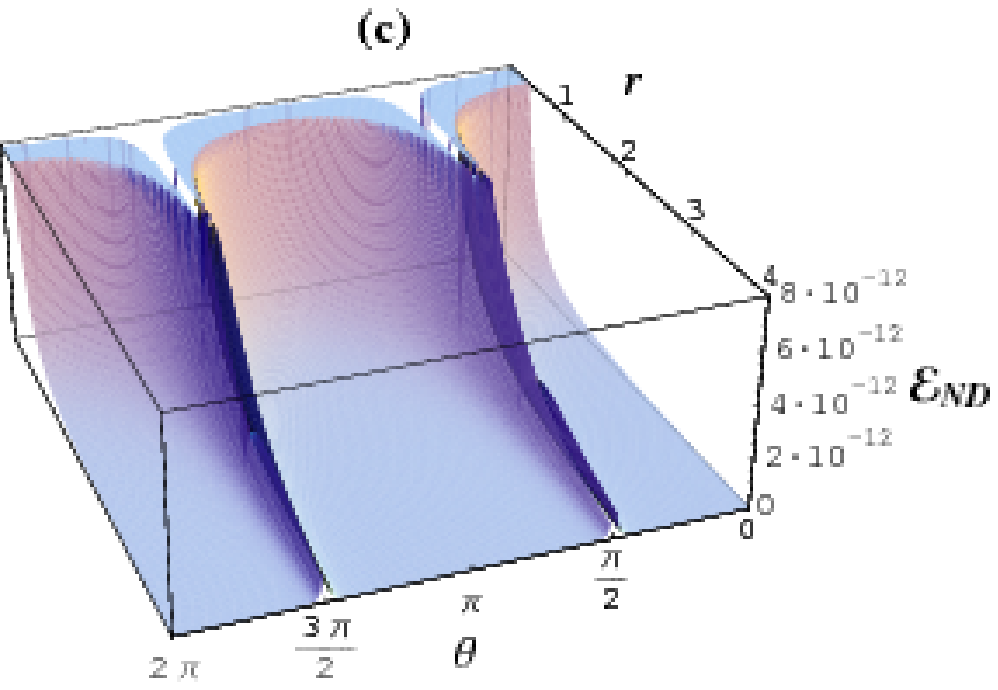}

\caption{Probability density (left plot), local entanglement $\mathcal{E}_{D}$ in the discarding ensemble (center plot) and local entanglement $\mathcal{E}_{ND}$ in the nondiscarding ensemble (right plot) for the relative wavefunction  $\varphi_{210}(r,\theta,\phi)$ of a hydrogen atom. All plots are for $a=b=0.1$. The cut-off points for plots of $\mathcal{E}_{D}$ and $\mathcal{E}_{ND}$ are determined  from $\epsilon_\mathrm{MAX}$ in equation~(\ref{eq:epsilonmax})  with $\sigma=0.1$; specifically, $\mathcal{E}_{ND}^\mathrm{MAX}=p_{ab}^\mathrm{MAX} h(\epsilon_\mathrm{MAX})$, where $h$ is defined in equation~(\ref{eq:vonNeuman}).
\label{fig:Hatomplots}}
\end{figure}

\section{Discussion and Conclusions}\label{sec:conclusions}
Our approach allows us to analyze the distribution of entanglement after imperfect local position measurements in any smooth bipartite pure state.  Equations (\ref{eq:ChighD}) and (\ref{eq:squaredCforS}) are our main results, allowing us to calculate the concurrence in terms of simple derivatives of the wavefunction.  Equation (\ref{eq:epsilonfortrasformedx}) allows us to express the entanglement in the same local region in terms of an arbitrary linear transformation of the coordinates, and (\ref{eq:epsilonTotal2}) treats the important case where the motion separates into center-of-mass and relative coordinates.

The three examples of exactly integrable systems that we have discussed show a number of common features. First, there is generic behavior near nodes in the wavefunction. There is an apparent divergence in the entanglement in the discarding ensemble for a fixed region size, but this does not mean that large amounts of entanglement can be extracted from the continuous-variable wavefunction once the system has been localized in this region.  Our expressions for entanglement are always true only in the limit of small region sizes, and their domain of validity shrinks as we approach a node; the discarding entanglement remains finite so long as we take care always to remain within this domain.  Furthermore, when we measure the locations of the particles we are unlikely to find them near a node in the wavefunction, so the probability factor in equation~(\ref{eq:nondiscdef}) further suppresses the non-discarding entanglement relative to the discarding entanglement.

As the size of the measurement regions increases, our approach starts to break down because more than two eigenvalues of the reduced density matrix become important. We have explicitly computed the extent of this breakdown, giving the lowest-order corrections to our main results in Appendix~\ref{app:a4b4correction}.

As pointed out in \S\ref{sec:relative-coordinates}, free-particle wavefunctions do not give rise to any local entanglement. We have shown how our entanglement expressions are transformed when moving to other coordinates (e.g. center-of-mass and relative coordinates); however, it is important to realize that the entanglement we quantify  is still between the original subsystems. The transformation is only done for the convenience of the calculations. 

Our results for the WKB wavefunctions and for the hydrogen atom suggest an intriguing link between the interaction force and the local  entanglement, but the exact details of the relationship and  its generality need  to be further explored.  We also note that in this paper we have only considered pure states; the application of our approach to the mixed states will be discussed in another paper. 

\appendix

\section{Corrections to the local entanglement after two-party preliminary measurements}\label{app:a4b4correction}

The third eigenvalue of Alice's reduced density matrix in the discarding ensemble when both parties make preliminary measurements can be found by making the following additional substitutions in equation~(\ref{eq:AliceOnlyThirdEigval}):
\begin{eqnarray}
\rho^{(A)}_{20}&=&{b\over p}[\rho_{2000}+{b^2\over 6}(\rho_{2020}+2\rho_{2011}+\rho_{2002})]+\mathrm{O}(b^5);\nonumber\\
\rho^{(A)}_{02}&=&{b\over p}[\rho_{0200}+{b^2\over 6}(\rho_{0220}+2\rho_{0211}+\rho_{0202})]+\mathrm{O}(b^5);\nonumber\\
\rho^{(A)}_{21}&=&{b\over p}[\rho_{2100}+{b^2\over 6}(\rho_{2120}+2\rho_{2111}+\rho_{2102})]+\mathrm{O}(b^5);\nonumber\\
\rho^{(A)}_{12}&=&{b\over p}[\rho_{1200}+{b^2\over 6}(\rho_{1220}+2\rho_{1211}+\rho_{1202})]+\mathrm{O}(b^5);\nonumber\\
\rho^{(A)}_{22}&=&{b\over 2p}[\rho_{2200}+{b^2\over 6}(\rho_{2220}+2\rho_{2211}+\rho_{2202})]+\mathrm{O}(b^5).\nonumber\\
\end{eqnarray}
This gives
\begin{equation}
%Give values
\lambda_3=\frac{\lambda_{3}^{nu} }{\lambda_{3}^{de} },
\end{equation}
where the denominator is 
\begin{eqnarray}
\lambda_{3}^{de} & = & 120(\rho_{0002}\rho_{0100}\rho_{1000}+2\rho_{0011}\rho_{0100}\rho_{1000}\nonumber\\
 &  &\quad+\rho_{0020}\rho_{0100}\rho_{1000}+\rho_{0000}\rho_{0102}\rho_{1000}\nonumber\\
 &  &\quad +2\rho_{0000}\rho_{0111}\rho_{1000}+\rho_{0000}\rho_{0120}\rho_{1000}\nonumber\\
 &  &\quad+\rho_{0000}\rho_{0100}\rho_{1002}+2\rho_{0000}\rho_{0100}\rho_{1011}\nonumber\\
 &  &\quad +\rho_{0000}\rho_{0100}\rho_{1020}-2\rho_{0000}\rho_{0002}\rho_{1100}\nonumber\\
 &  &\quad-4\rho_{0000}\rho_{0011}\rho_{1100}-2\rho_{0000}\rho_{0020}\rho_{1100}\nonumber\\
 &  &\quad -\rho_{0000}\rho_{0000}\rho_{1102}-2\rho_{0000}\rho_{0000}\rho_{1111}\nonumber\\
 &  &\quad-\rho_{0000}\rho_{0000}\rho_{1120}),\end{eqnarray}
and the numerator is
\begin{eqnarray}
\lambda_{3}^{nu} & = &\frac{1}{54} (\rho_{0211}\rho_{1120}\rho_{2002}-\rho_{0120}\rho_{1211}\rho_{2002}\nonumber\\
 &  & \quad-\rho_{0111}\rho_{1220}\rho_{2002}+\rho_{0220}\rho_{1111}\rho_{2002}\nonumber\\
 &  &\quad +\rho_{0202}\rho_{1120}\rho_{2011}-\rho_{0120}\rho_{1202}\rho_{2011}\nonumber\\
 &  & \quad-\rho_{0102}\rho_{1220}\rho_{2011}+\rho_{0220}\rho_{1102}\rho_{2011}\nonumber\\
 &  & \quad-\rho_{0220}\rho_{1011}\rho_{2102}-\rho_{0211}\rho_{1020}\rho_{2102}\nonumber\\
 &  & \quad+\rho_{0020}\rho_{1211}\rho_{2102}+\rho_{0011}\rho_{1220}\rho_{2102}\nonumber\\
 &  & \quad-\rho_{0220}\rho_{1002}\rho_{2111}-\rho_{0202}\rho_{1020}\rho_{2111}\nonumber\\
 &  & \quad+\rho_{0020}\rho_{1202}\rho_{2111}+\rho_{0002}\rho_{1220}\rho_{2111}\nonumber\\
 &  & \quad-\rho_{0211}\rho_{1002}\rho_{2120}+\rho_{0002}\rho_{1211}\rho_{2120}\nonumber\\
 &  & \quad+\rho_{0211}\rho_{1000}\rho_{2122}-\rho_{0000}\rho_{1211}\rho_{2122}\nonumber\\
 &  &\quad +\rho_{0120}\rho_{1011}\rho_{2202}+\rho_{0111}\rho_{1020}\rho_{2202}\nonumber\\
 &  &\quad -\rho_{0020}\rho_{1111}\rho_{2202}-\rho_{0011}\rho_{1120}\rho_{2202}\nonumber\\
 &  & \quad+\rho_{0120}\rho_{1002}\rho_{2211}+\rho_{0102}\rho_{1020}\rho_{2211}\nonumber\\
 &  & \quad-\rho_{0020}\rho_{1102}\rho_{2211}-\rho_{0002}\rho_{1120}\rho_{2211}\nonumber\\
 &  & \quad+\rho_{0111}\rho_{1002}\rho_{2220}+\rho_{0102}\rho_{1011}\rho_{2220}\nonumber\\
 &  & \quad-\rho_{0011}\rho_{1102}\rho_{2220}-\rho_{0002}\rho_{1111}\rho_{2220}\nonumber\\
 &  & \quad-\rho_{0111}\rho_{1000}\rho_{2222}-\rho_{0100}\rho_{1011}\rho_{2222}\nonumber\\
 &  & \quad+\rho_{0011}\rho_{1100}\rho_{2222}+\rho_{0000}\rho_{1111}\rho_{2222}).\end{eqnarray}

%\section*{References}
\bibliographystyle{apsrev} \bibliographystyle{apsrev}
\bibliography{multiD_pure_reference}

\begin{thebibliography}{24}
\expandafter\ifx\csname natexlab\endcsname\relax\def\natexlab#1{#1}\fi
\expandafter\ifx\csname bibnamefont\endcsname\relax
  \def\bibnamefont#1{#1}\fi
\expandafter\ifx\csname bibfnamefont\endcsname\relax
  \def\bibfnamefont#1{#1}\fi
\expandafter\ifx\csname citenamefont\endcsname\relax
  \def\citenamefont#1{#1}\fi
\expandafter\ifx\csname url\endcsname\relax
  \def\url#1{\texttt{#1}}\fi
\expandafter\ifx\csname urlprefix\endcsname\relax\def\urlprefix{URL }\fi
\providecommand{\bibinfo}[2]{#2}
\providecommand{\eprint}[2][]{\url{#2}}

\bibitem[{\citenamefont{Nielsen and Chuang}(2000)}]{Nielsenbook}
\bibinfo{author}{\bibfnamefont{M.~A.} \bibnamefont{Nielsen}} \bibnamefont{and}
  \bibinfo{author}{\bibfnamefont{I.}~\bibnamefont{Chuang}},
  \emph{\bibinfo{title}{Quantum Computation and Quantum Information}}
  (\bibinfo{publisher}{Cambridge University Press},
  \bibinfo{address}{Cambridge}, \bibinfo{year}{2000}).

\bibitem[{\citenamefont{Ekert}(1991)}]{Ekert1991}
\bibinfo{author}{\bibfnamefont{A.~K.} \bibnamefont{Ekert}},
  \bibinfo{journal}{Phys.\ Rev.\ Lett.} \textbf{\bibinfo{volume}{67}},
  \bibinfo{pages}{661} (\bibinfo{year}{1991}).

\bibitem[{\citenamefont{Bennett et~al.}(1993)\citenamefont{Bennett, Brassard,
  Crepeaua, Josza, Peres, and Wootters}}]{Bennett1993}
\bibinfo{author}{\bibfnamefont{C.~H.} \bibnamefont{Bennett}},
  \bibinfo{author}{\bibfnamefont{G.}~\bibnamefont{Brassard}},
  \bibinfo{author}{\bibfnamefont{C.}~\bibnamefont{Crepeaua}},
  \bibinfo{author}{\bibfnamefont{R.}~\bibnamefont{Josza}},
  \bibinfo{author}{\bibfnamefont{A.}~\bibnamefont{Peres}}, \bibnamefont{and}
  \bibinfo{author}{\bibfnamefont{W.~K.} \bibnamefont{Wootters}},
  \bibinfo{journal}{Phys.\ Rev.\ Lett.} \textbf{\bibinfo{volume}{70}},
  \bibinfo{pages}{1895} (\bibinfo{year}{1993}).

\bibitem[{\citenamefont{Braunstein and van
  Loock}(2005)}]{Braunstein.RevModPhys.77.531}
\bibinfo{author}{\bibfnamefont{S.~L.} \bibnamefont{Braunstein}}
  \bibnamefont{and} \bibinfo{author}{\bibfnamefont{P.}~\bibnamefont{van
  Loock}}, \bibinfo{journal}{Rev. Mod. Phys.} \textbf{\bibinfo{volume}{77}},
  \bibinfo{pages}{531} (\bibinfo{year}{2005}).

\bibitem[{\citenamefont{Braunstein and Pati}(2003)}]{Braunsteinbook}
\bibinfo{author}{\bibfnamefont{S.~L.} \bibnamefont{Braunstein}}
  \bibnamefont{and} \bibinfo{author}{\bibfnamefont{A.~K.} \bibnamefont{Pati}},
  \emph{\bibinfo{title}{Quantum Information Theory with Continuous Variables}}
  (\bibinfo{publisher}{Kluwer Academic Press}, \bibinfo{address}{Dordrecht},
  \bibinfo{year}{2003}).

\bibitem[{\citenamefont{Eisert and Plenio}(2003)}]{Eisert2003}
\bibinfo{author}{\bibfnamefont{J.}~\bibnamefont{Eisert}} \bibnamefont{and}
  \bibinfo{author}{\bibfnamefont{M.~B.} \bibnamefont{Plenio}},
  \bibinfo{journal}{Int. J. Quant. Inf.} \textbf{\bibinfo{volume}{1}},
  \bibinfo{pages}{479} (\bibinfo{year}{2003}).

\bibitem[{\citenamefont{Duan et~al.}(2000)\citenamefont{Duan, G.~Giedke, and
  Zoller}}]{Duan2000}
\bibinfo{author}{\bibfnamefont{L.-M.} \bibnamefont{Duan}},
  \bibinfo{author}{\bibfnamefont{J.~I.~C.} \bibnamefont{G.~Giedke}},
  \bibnamefont{and} \bibinfo{author}{\bibfnamefont{P.}~\bibnamefont{Zoller}},
  \bibinfo{journal}{Phys. Rev. Lett.} \textbf{\bibinfo{volume}{84}},
  \bibinfo{pages}{4002} (\bibinfo{year}{2000}).

\bibitem[{\citenamefont{Eisert et~al.}(2002)\citenamefont{Eisert, Scheel, and
  Plenio}}]{Eisert2002}
\bibinfo{author}{\bibfnamefont{J.}~\bibnamefont{Eisert}},
  \bibinfo{author}{\bibfnamefont{S.}~\bibnamefont{Scheel}}, \bibnamefont{and}
  \bibinfo{author}{\bibfnamefont{M.~B.} \bibnamefont{Plenio}},
  \bibinfo{journal}{Phys. Rev. Lett.} \textbf{\bibinfo{volume}{89}},
  \bibinfo{pages}{137903} (\bibinfo{year}{2002}).

\bibitem[{\citenamefont{Fiurasek}(2002)}]{Fiurasek2002}
\bibinfo{author}{\bibfnamefont{J.}~\bibnamefont{Fiurasek}},
  \bibinfo{journal}{Phys. Rev. Lett.} \textbf{\bibinfo{volume}{89}},
  \bibinfo{pages}{137904} (\bibinfo{year}{2002}).

\bibitem[{\citenamefont{Giedke and Cirac}(2002)}]{Giedke2002}
\bibinfo{author}{\bibfnamefont{G.}~\bibnamefont{Giedke}} \bibnamefont{and}
  \bibinfo{author}{\bibfnamefont{J.~I.} \bibnamefont{Cirac}},
  \bibinfo{journal}{Phys. Rev. A} \textbf{\bibinfo{volume}{66}},
  \bibinfo{pages}{0323} (\bibinfo{year}{2002}).

\bibitem[{\citenamefont{van Loock and Braunstein}(2000)}]{Loock2000}
\bibinfo{author}{\bibfnamefont{P.}~\bibnamefont{van Loock}} \bibnamefont{and}
  \bibinfo{author}{\bibfnamefont{S.~L.} \bibnamefont{Braunstein}},
  \bibinfo{journal}{Phys. Rev. Lett.} \textbf{\bibinfo{volume}{84}},
  \bibinfo{pages}{3482} (\bibinfo{year}{2000}).

\bibitem[{\citenamefont{Giedke et~al.}(2001)\citenamefont{Giedke, Kraus,
  Lewenstein, and Cirac}}]{Giedke2001}
\bibinfo{author}{\bibfnamefont{G.}~\bibnamefont{Giedke}},
  \bibinfo{author}{\bibfnamefont{B.}~\bibnamefont{Kraus}},
  \bibinfo{author}{\bibfnamefont{M.}~\bibnamefont{Lewenstein}},
  \bibnamefont{and} \bibinfo{author}{\bibfnamefont{J.~I.} \bibnamefont{Cirac}},
  \bibinfo{journal}{Phys. Rev. A} \textbf{\bibinfo{volume}{64}},
  \bibinfo{pages}{052303} (\bibinfo{year}{2001}).

\bibitem[{\citenamefont{Adesso et~al.}(2004)\citenamefont{Adesso, Seraffin, and
  Illuminati}}]{Adesso2004}
\bibinfo{author}{\bibfnamefont{G.}~\bibnamefont{Adesso}},
  \bibinfo{author}{\bibfnamefont{A.}~\bibnamefont{Seraffin}}, \bibnamefont{and}
  \bibinfo{author}{\bibfnamefont{F.}~\bibnamefont{Illuminati}},
  \bibinfo{journal}{Phys. Rev. Lett.} \textbf{\bibinfo{volume}{93}},
  \bibinfo{pages}{220504} (\bibinfo{year}{2004}).

\bibitem[{\citenamefont{Giedke et~al.}(2003)\citenamefont{Giedke, Wolf, Kruger,
  Werner, and Cirac}}]{giedke:107901}
\bibinfo{author}{\bibfnamefont{G.}~\bibnamefont{Giedke}},
  \bibinfo{author}{\bibfnamefont{M.~M.} \bibnamefont{Wolf}},
  \bibinfo{author}{\bibfnamefont{O.}~\bibnamefont{Kruger}},
  \bibinfo{author}{\bibfnamefont{R.~F.} \bibnamefont{Werner}},
  \bibnamefont{and} \bibinfo{author}{\bibfnamefont{J.~I.} \bibnamefont{Cirac}},
  \bibinfo{journal}{Phys. Rev. Lett.} \textbf{\bibinfo{volume}{91}},
  \bibinfo{pages}{107901} (\bibinfo{year}{2003}).

\bibitem[{\citenamefont{Wolf et~al.}(2004)\citenamefont{Wolf, Giedke, Kruger,
  Werner, and Cirac}}]{Wolf2004}
\bibinfo{author}{\bibfnamefont{M.~M.} \bibnamefont{Wolf}},
  \bibinfo{author}{\bibfnamefont{G.}~\bibnamefont{Giedke}},
  \bibinfo{author}{\bibfnamefont{O.}~\bibnamefont{Kruger}},
  \bibinfo{author}{\bibfnamefont{R.~F.} \bibnamefont{Werner}},
  \bibnamefont{and} \bibinfo{author}{\bibfnamefont{J.~I.} \bibnamefont{Cirac}},
  \bibinfo{journal}{Phys. Rev. A} \textbf{\bibinfo{volume}{69}},
  \bibinfo{pages}{052320} (\bibinfo{year}{2004}).

\bibitem[{\citenamefont{Vidal and Werner}(2002)}]{PhysRevA.65.032314}
\bibinfo{author}{\bibfnamefont{G.}~\bibnamefont{Vidal}} \bibnamefont{and}
  \bibinfo{author}{\bibfnamefont{R.~F.} \bibnamefont{Werner}},
  \bibinfo{journal}{Phys. Rev. A} \textbf{\bibinfo{volume}{65}},
  \bibinfo{pages}{032314} (\bibinfo{year}{2002}).

\bibitem[{\citenamefont{Audenaert et~al.}(2002)\citenamefont{Audenaert, Eisert,
  Plenio, and Werner}}]{PhysRevA.66.042327}
\bibinfo{author}{\bibfnamefont{K.}~\bibnamefont{Audenaert}},
  \bibinfo{author}{\bibfnamefont{J.}~\bibnamefont{Eisert}},
  \bibinfo{author}{\bibfnamefont{M.~B.} \bibnamefont{Plenio}},
  \bibnamefont{and} \bibinfo{author}{\bibfnamefont{R.~F.}
  \bibnamefont{Werner}}, \bibinfo{journal}{Phys. Rev. A}
  \textbf{\bibinfo{volume}{66}}, \bibinfo{pages}{042327}
  (\bibinfo{year}{2002}).

\bibitem[{\citenamefont{Shchukin and Vogel}(2005)}]{shchukin05}
\bibinfo{author}{\bibfnamefont{E.}~\bibnamefont{Shchukin}} \bibnamefont{and}
  \bibinfo{author}{\bibfnamefont{W.}~\bibnamefont{Vogel}},
  \bibinfo{journal}{Phys.\ Rev.\ Lett.} \textbf{\bibinfo{volume}{95}},
  \bibinfo{pages}{230502} (\bibinfo{year}{2005}).

\bibitem[{\citenamefont{Lin and Fisher}(2007{\natexlab{a}})}]{lin1}
\bibinfo{author}{\bibfnamefont{H.-C.} \bibnamefont{Lin}} \bibnamefont{and}
  \bibinfo{author}{\bibfnamefont{A.~J.} \bibnamefont{Fisher}},
  \bibinfo{journal}{Phys.\ Rev.\ A} \textbf{\bibinfo{volume}{75}},
  \bibinfo{pages}{032330} (\bibinfo{year}{2007}{\natexlab{a}}).

\bibitem[{\citenamefont{Lin and Fisher}(2007{\natexlab{b}})}]{lin2}
\bibinfo{author}{\bibfnamefont{H.-C.} \bibnamefont{Lin}} \bibnamefont{and}
  \bibinfo{author}{\bibfnamefont{A.~J.} \bibnamefont{Fisher}},
  \bibinfo{journal}{Phys.\ Rev.\ A} \textbf{\bibinfo{volume}{76}},
  \bibinfo{pages}{042320} (\bibinfo{year}{2007}{\natexlab{b}}).

\bibitem[{\citenamefont{Wootters}(1998)}]{wootters98}
\bibinfo{author}{\bibfnamefont{W.~K.} \bibnamefont{Wootters}},
  \bibinfo{journal}{Phys. Rev. Lett.} \textbf{\bibinfo{volume}{80}},
  \bibinfo{pages}{2245} (\bibinfo{year}{1998}).

\bibitem[{\citenamefont{Eisert and Plenio}(1999)}]{JModOptic46.145}
\bibinfo{author}{\bibfnamefont{J.}~\bibnamefont{Eisert}} \bibnamefont{and}
  \bibinfo{author}{\bibfnamefont{M.~B.} \bibnamefont{Plenio}},
  \bibinfo{journal}{J. Mod. Optic.} \textbf{\bibinfo{volume}{46}},
  \bibinfo{pages}{145} (\bibinfo{year}{1999}).

\bibitem[{\citenamefont{Zyczkowski et~al.}(1998)\citenamefont{Zyczkowski,
  Horodecki, Sanpera, and Lewenstein}}]{Zyczkowski1998}
\bibinfo{author}{\bibfnamefont{K.}~\bibnamefont{Zyczkowski}},
  \bibinfo{author}{\bibfnamefont{P.}~\bibnamefont{Horodecki}},
  \bibinfo{author}{\bibfnamefont{A.}~\bibnamefont{Sanpera}}, \bibnamefont{and}
  \bibinfo{author}{\bibfnamefont{M.}~\bibnamefont{Lewenstein}},
  \bibinfo{journal}{Phys. Rev. A} \textbf{\bibinfo{volume}{58}},
  \bibinfo{pages}{883} (\bibinfo{year}{1998}).

\bibitem[{\citenamefont{Chen et~al.}(2005)\citenamefont{Chen, Albeverio, and
  Fei}}]{PRL95.040504}
\bibinfo{author}{\bibfnamefont{K.}~\bibnamefont{Chen}},
  \bibinfo{author}{\bibfnamefont{S.}~\bibnamefont{Albeverio}},
  \bibnamefont{and} \bibinfo{author}{\bibfnamefont{S.-M.} \bibnamefont{Fei}},
  \bibinfo{journal}{Phys. Rev. Lett.} \textbf{\bibinfo{volume}{95}}
  (\bibinfo{year}{2005}).

\end{thebibliography}

\end{document}